\pdfoutput=1

\documentclass[9pt]{article}

\usepackage{geometry}
\usepackage[utf8]{inputenc}   
\usepackage[T1]{fontenc}      
\usepackage{hyperref}         
\usepackage{url}              
\usepackage{booktabs}         
\usepackage{amsfonts}         
\usepackage{nicefrac}         
\usepackage{microtype}        
\usepackage{xcolor}           
\usepackage[numbers]{natbib}
\usepackage{enumitem}         
\usepackage{caption,bm}       
\usepackage{subfigure}        
\usepackage{graphicx}         
\usepackage{diagbox,multirow} 
\usepackage{algorithm}        
\usepackage{algorithmic}      
\usepackage{amsmath,amssymb,bbm} 

\usepackage{amsthm}           

\usepackage[english]{babel}

\usepackage{pifont}           
\usepackage{subcaption}       
\usepackage{mathtools}        
\usepackage{wrapfig}          
\usepackage{titletoc}         
\newcommand\DoToC{%
  \startcontents
  \printcontents{}{1}{\vskip3pt\hrule\vskip5pt}
  \vskip15pt\hrule\vskip5pt
}

\newcommand{\cmark}{\textcolor{blue}{\ding{51}}}%
\newcommand{\xmark}{\textcolor{red}{\ding{55}}}%
\newcommand{\proposed}{\textsf{PCRL}}

\newcommand{\fullname}{\textsf{P}robabilistic \textsf{C}ompositional \textsf{R}epresentation \textsf{L}earning}

\usepackage{authblk}

\title{Compositional Representation of Polymorphic Crystalline Materials}

\author[1]{Namkyeong Lee}
\author[1]{Heewoong Noh}
\author[2]{Gyoung S. Na}
\author[3]{Jimeng Sun}
\author[4]{Tianfan Fu}
\author[5]{Marinka Zitnik}
\author[1]{Chanyoung Park}

\affil[1]{Korea Advanced Institute of Science and Technology (KAIST)}
\affil[2]{Korea Research Institute of Chemical Technology (KRICT)}
\affil[3]{University of Illinois Urbana-Champaign (UIUC)}
\affil[4]{Rensselaer Polytechnic Institute (RPI)}
\affil[5]{Harvard University}

\begin{document}

\maketitle

\begin{abstract}
Machine learning (ML) has seen promising developments in materials science, yet its efficacy largely depends on detailed crystal structural data, which are often complex and hard to obtain, limiting their applicability in real-world material synthesis processes. 
An alternative, using compositional descriptors, offers a simpler approach by indicating the elemental ratios of compounds without detailed structural insights. 
However, accurately representing materials solely with compositional descriptors presents challenges due to polymorphism, where a single composition can correspond to various structural arrangements, creating ambiguities in its representation.
To this end, we introduce \proposed, a novel approach that employs probabilistic modeling of composition to capture the diverse polymorphs from available structural information.
Extensive evaluations on sixteen datasets demonstrate the effectiveness of \proposed~in learning compositional representation, and our analysis highlights its potential applicability of \proposed~in material discovery.
The source code for \proposed~is available at~\url{https://github.com/Namkyeong/PCRL}.
\end{abstract}

\section{Introduction}

\begin{figure*}[t!]
    \centering
    \includegraphics[width=0.99\linewidth]{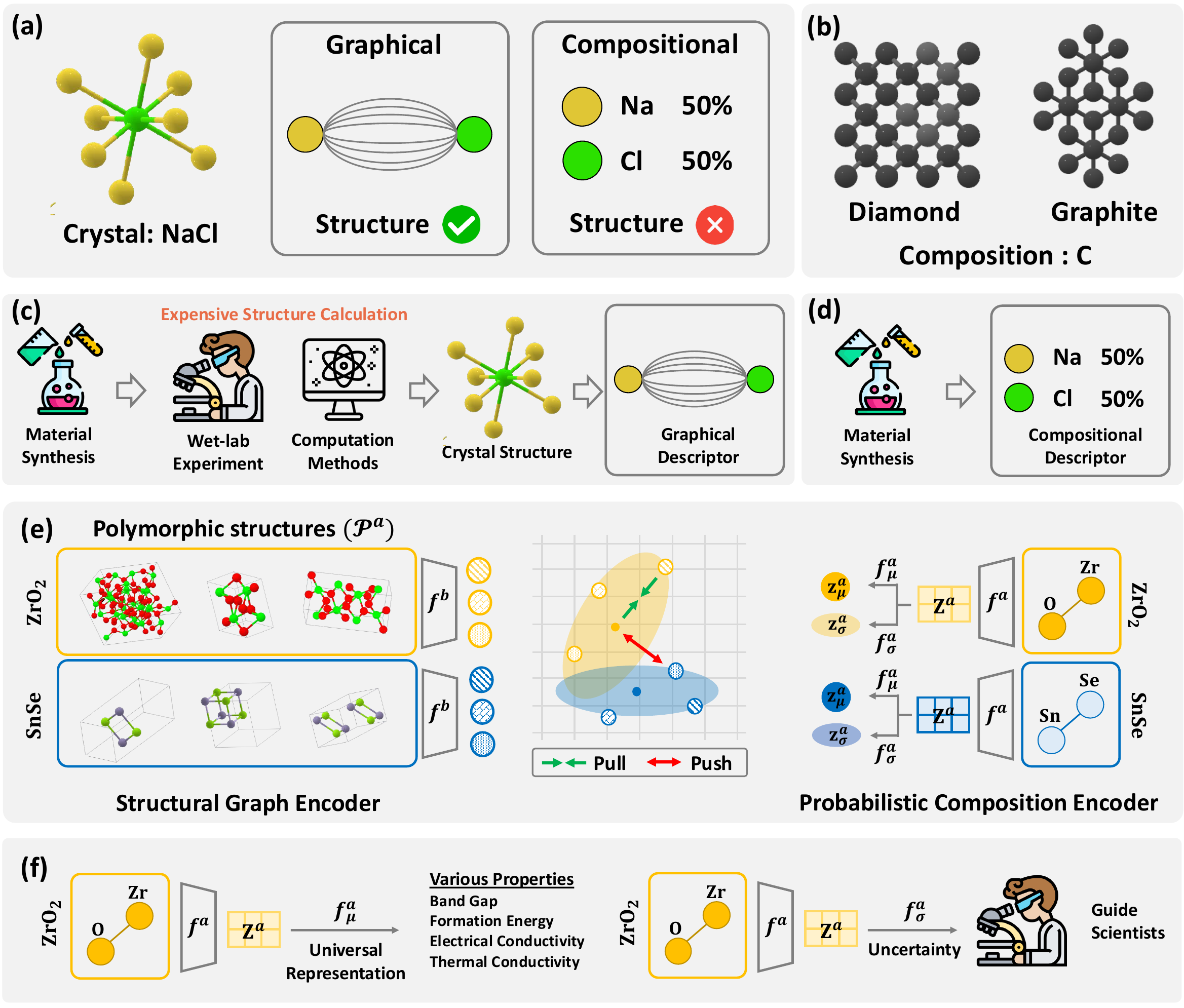}
    \caption{
    (a) A crystal can have multiple descriptors for ML input, such as graphical and compositional descriptors.
    (b) Diamond and Graphite are polymorphic crystal structures of composition C. While these crystal structures share the same composition, their properties are completely different.
    (c) Graphical descriptors necessitate costly recalculation of crystal structures during the material synthesis process, thereby restricting the ML capability to the same bottleneck as conventional materials discovery processes. 
    (d) Compositional representation enables the use of ML without requiring these expensive structural calculations.
    (e) Pre-training composition graph encoder encoder with contrastive learning. While the structural graph encoder obtains a deterministic structural representation of crystal, the probabilistic composition encoder learns to represent each composition as a parameterized probabilistic distribution by acquiring mean and diagonal covariance matrices. Both encoders are jointly trained with soft contrastive loss in representation space.
    (f) The pre-trained composition mean encoder $f_{\mu}^{a}$ can be utilized to predict various properties of materials, while the composition uncertainty encoder $f_{\sigma}^{a}$ can guide scientists in determining which materials to investigate further.}
    \label{fig1}
\end{figure*}


Recently, machine learning (ML) has increasingly been leveraged in materials science to sift through the abundant experimental and computational data available~\citep{zhang2023artificial,wang2023scientific}. 
Interest has particularly spiked in graphical descriptors derived from the structural properties of crystals. 
Notably, graphical representations can be created by incorporating periodic boundary conditions and connecting neighboring atoms within a certain range, as depicted in Figure~\ref{fig1}~(a) \citep{xie2018crystal,chen2019graph}. 
However, in the real-world material synthesis process, the reorganization of atomic arrangements frequently occurs due to the mixing of ingredients, heat treatments, and chemical reactions with solvents, necessitating the recalculation of crystal structures.
These recalculations rely on the computationally intensive and sometimes infeasible Density Functional Theory (DFT)~\citep{sholl2022density}, limiting the ML capability to the same computational bottleneck as DFT calculations.
In other words, while ML has garnered significant interest from material scientists because of its efficiency and speed compared to DFT calculations, it still necessitates DFT calculations when utilizing graphical representations of materials in ML models during the materials discovery process.
Moreover, in materials science, doping and alloying are used to improve the performance of materials such as semiconductors ~\citep{kawai1992effects,jin2014doping}.
However, because doping and alloying chaotically change the structure of pure materials, it can lead to cases where calculations of the structure are infeasible, limiting the use of graphical representation for materials.

An alternative to graphical descriptors is to utilize material representations based purely on \textit{composition}, which denotes the elemental ratios present in a chemical compound, as illustrated in Figure~\ref{fig1}~(a)~\citep{jha2018elemnet,goodall2020predicting}. 
While straightforward, models based on composition have proven to be effective, offering a promising array of potential elemental combinations for new material discovery at a low computational expense\citep{damewood2023representations}. 
Nonetheless, this method has a significant drawback as it ignores the structural aspects of crystals, which can result in less effective performance compared to graphical models~\citep{bartel2020critical}. 
This discrepancy emphasizes the structural impact on crystal properties and raises an important inquiry: ``\textit{Can composition-based models be enhanced by incorporating structural details of crystals?}''

Recent multi-modal contrastive pre-training strategies in various domains may provide an answer to this question \citep{gan2022vision, zong2023self}. 
Specifically, the CLIP model \citep{radford2021learning} in computer vision significantly enhances zero-shot learning by aligning images with their captions. 
Moreover, 3D Infomax \citep{stark20223d} improves the accuracy of 2D molecular graphs in quantum chemistry by maximizing their mutual information with 3D molecules during the pre-training step and then using only 2D molecular graphs during the fine-tuning step.
Similarly, one could introduce both compositional and structural information as multiple modalities for materials during the pre-training stage, and then use only the compositional information during the fine-tuning stage.

However, implementing multi-modal contrastive learning techniques directly for compositional representation learning encounters obstacles. 
This is due to the inherent nature of crystal structures, named polymorphism, where a single composition can exist in multiple crystal structures.
That is, the relationship between composition and crystal structures is one-to-many rather than one-to-one, leading to intrinsic uncertainties in practical applications.

More specifically, polymorphism describes how a single compound can form various crystallographic structures due to different atomic arrangements, which leads to vastly different physical and chemical properties~\citep{bernstein2020polymorphism}. 
For example, carbon can exist as either diamond or graphite as shown in Figure \ref{fig1} (b). 
In diamond, carbon atoms are arranged in a tetrahedral lattice with each carbon atom bonded to four others, giving it extraordinary hardness and unique optical properties~\citep{che2000thermal,kidalov2009thermal}. 
In contrast, graphite has a planar layered structure where carbon atoms are bonded in hexagonal rings, allowing the sheets to slide over each other easily, which results in its lubricating and electrical conductivity properties~\citep{wissler2006graphite,jorio2008carbon}.
Although the compositional models give up the ability to handle such polymorphs while circumventing the structure bottleneck \citep{goodall2020predicting}, one can alternatively provide uncertainties caused by polymorphism, which is crucial in real-world materials discovery.


\subsection{Our Approach}
To this end, we propose a multi-modal pre-training framework for crystalline materials, namely \textsf{P}robabilistic \textsf{C}ompositional \textsf{R}epresentation \textsf{L}earning (\proposed).
This framework aims to pre-train a composition encoder that can be universally applied to various downstream tasks.
Specifically, the composition encoder outputs representation of composition as a probabilistic distribution of polymorphs instead of a single deterministic representation~\citep{oh2018modeling,chun2021probabilistic}.
In particular, by assuming that polymorphs with an identical composition follow the same Gaussian distribution, \proposed~models each composition as a parameterized Gaussian distribution with learnable mean and variance vectors, whose distribution is trained to cover the range of polymorphic structures in representation space.
By doing so, we expect the mean of the parameterized Gaussian distribution to represent the composition and the variance to reflect the uncertainty stemming from the existence of various polymorphic structures.

In this work, we make the following contributions:
\begin{itemize}[leftmargin=.1in]
    \item Recognizing the advantages and limitations of both structural and compositional descriptors, we propose a multi-modal representation learning framework for composition, called~\proposed, which incorporates structural information of crystals into compositional representations.
    \item Unlike previous material property prediction works that rely on composition, our goal is to pre-train a universally applicable composition encoder that can be used for predicting various properties.
    \item To capture uncertainties of composition stemming from various \textit{polymorphs}, \proposed~learns a probabilistic representation for each composition, whose mean serves as the representation of the composition and variance reflects uncertainty stemming from the polymorphism.
    \item Extensive experiments on \textbf{sixteen datasets} demonstrate the superiority of \proposed~in learning universally applicable representation of composition and predicting its various physical properties. Moreover, measured uncertainties reflect various challenges in materials science, highlighting the applicability of \proposed~for real-world materials discovery.
\end{itemize}
To the best of our knowledge, this is the \textbf{first work that learns universal compositional representations} by simultaneously considering the crystal structural information and the polymorphism as uncertainty, which is crucial for real-world material discovery.

\section{Method}

\subsection{Preliminaries}
\subsubsection{Compositional Graph Construction}
\label{sec: compositional Graph Construction}
Given a composition, we use $\mathcal{E} = \{e_1, \ldots, e_{n_e} \}$ to denote its unique set of elements, and $\mathcal{R} = \{r_1, \ldots, r_{n_e} \}$ to denote the compositional ratio of each element in the composition.
We construct a fully connected compositional graph $\mathcal{G}^a = (\mathcal{E}, \mathcal{R}, \mathbf{A}^a)$, where $\mathbf{A}^a \in \{1\}^{{n_e} \times {n_e}}$ indicates the adjacency matrix of a fully connected graph~\citep{goodall2020predicting}.
Then, we adopt GNNs as the composition encoder $f^a$, which aims to learn the compositional representation by capturing complex relationships between elements via the message-passing scheme.
Additionally, $\mathcal{G}^a$ is associated with an elemental feature matrix $\mathbf{X}^a \in \mathbb{R}^{{n_e} \times F}$ where $F$ is the number of features.

\subsubsection{Structural Graph Construction}
\label{sec: Structural Graph Construction}
Given a crystal structure $(\mathbf{P}, \mathbf{L})$, suppose the unit cell has $n_s$ atoms, we have $\mathbf{P} = {[\mathbf{p}_1, \mathbf{p}_2, \ldots, \mathbf{p}_{n_s}]}^\intercal \in \mathbb{R}^{{n_s} \times 3}$ indicating the atom position matrix and $\mathbf{L} = {[\mathbf{l}_1, \mathbf{l}_2, \mathbf{l}_3]}^\intercal \in \mathbb{R}^{3 \times 3}$ representing the lattice parameter describing how a unit cell repeats itself in three directions.
Based on the crystal parameters, we construct a multi-edge graph $\mathcal{G}^b = (\mathcal{V}, \mathbf{A}^b)$ that captures atom interactions across cell boundaries~\citep{xie2018crystal}.
Specifically, $v_i \in \mathcal{V}$ denotes an atom $i$ and all its duplicates in the infinite 3D space whose positions are included in the set $\{\hat{\mathbf{p}}_i | \hat{\mathbf{p}}_i = \mathbf{p}_i+k_1\mathbf{l}_1+k_2\mathbf{l}_2+k_3\mathbf{l}_3, k_1, k_2, k_3 \in \mathbb{Z} \}$, where $\mathbb{Z}$ denotes the set of all the integers.
Moreover, $\mathbf{A}^b \in \{0,1\}^{n_s\times n_s}$ denotes an adjacency matrix, where $\mathbf{A}_{i,j}^b = 1$ if two atoms $i$ and $j$ are within the predefined radius $r$ and $\mathbf{A}^{b}_{ij} = 0$ otherwise.
Furthermore, a single compositional graph $\mathcal{G}^a$ is associated with a set of polymorphic crystal structural graphs $\mathcal{P}^{\mathcal{G}^a}$, i.e., $\mathcal{P}^{\mathcal{G}^a} = \{\mathcal{G}_1^b, \ldots, \mathcal{G}_{n_p}^b \}$, where $n_p$ is the number of polymorphs for the composition.
We provide further details on structural graph construction in Appendix~\ref{App: Structural Graph Construction}.

\subsubsection{Task Description}
Given the compositional graph $\mathcal{G}^a$ and the structural graph $\mathcal{G}^b$ of a single crystal, our objective is to acquire a composition encoder denoted as $f^a$, alongside mean and variance modules referred to as $f_{\mu}^a$ and $f_{\sigma}^a$.
Then, these modules including pre-trained composition encoder $f^a$, and the mean $f_{\mu}^a$ and variance $f_{\sigma}^a$ modules, are employed in a range of downstream tasks, a scenario frequently encountered in real-world material synthesizing processes where \textit{solely composition of crystalline material is accessible}.


\subsection{Structural Graph Encoder $f^b$}\label{sec:sge}
While structural information plays an important role in determining various properties of crystals, 
previous studies have overlooked the readily available crystal structures~\citep{jain2013commentary} for compositional representation learning~\citep{jha2018elemnet, goodall2020predicting, wang2021compositionally}.
To this end, we use a GNN encoder to learn the representation of crystal structure, which is expected to provide guidance for learning the representation of composition.
More formally, given the crystal structural graph $\mathcal{G}^{b} = (\mathbf{x}^b, \mathbf{A}^b)$, we obtain a structural representation of a crystal as follows:
\begin{equation}
\small
    \mathbf{z}^b = \text{Pooling}(\mathbf{Z}^b),\,\,\,\,  \mathbf{Z}^b = f^b (\mathbf{x}^b, \mathbf{A}^b),
    \label{eq: structural}
\end{equation}
where $\mathbf{Z}^b \in \mathbb{R}^{n_s \times F}$ is a matrix whose each row indicates the representation of each atom in the crystal structure, $\mathbf{z}^b$ indicates the latent representation of a crystal structure, and $f^b$ is the GNN-based crystal structural encoder.
In this paper, we adopt graph neural networks~\citep{battaglia2018relational} as the encoder, which is a generalized version of various GNNs, and sum pooling is used as the pooling function.
We provide further details on the GNNs in Appendix~\ref{App: Structural Graph Encoder}.

\subsection{Probabilistic Composition Encoder $f^a$}\label{sec:pse}
\textbf{Deterministic Representation.} After obtaining the structural representation $\mathbf{z}^b$, we also compute the compositional representation from the composition graph $\mathcal{G}^a$ as follows:
\begin{equation}
\small
    \mathbf{z}^a = \text{Pooling}(\mathbf{Z}^a),\,\,\,\, \mathbf{Z}^a = f^a (\mathbf{X}^a, \mathbf{A}^a), 
    \label{eq: composition}
\end{equation}
where $\mathbf{Z}^a \in \mathbb{R}^{n_e \times F}$ is a matrix whose each row indicates the representation of each element in a composition, $\mathbf{z}^a \in \mathbb{R}^{F}$ indicates the compositional representation of a crystal, and $f^a$ is a GNN-based composition encoder.
By utilizing GNNs, the composition encoder effectively learns intricate relationships and chemical environments related to elements, thereby enhancing the compositional representation in a systematic manner~\citep{goodall2020predicting}.
For the composition encoder $f^a$, we adopt GCNs~\citep{kipf2016semi} with jumping knowledge~\citep{xu2018representation},
and weighted sum pooling with the compositional ratio (i.e., $\mathcal{R}$ in Section~\ref{sec: compositional Graph Construction}) is used as the pooling function.

One straightforward approach for injecting structural information into the compositional representation would be adopting the idea of recent multi-modal contrastive learning approaches, which have been widely known to maximize the mutual information between heterogeneous modality inputs  (two modalities in our case: composition and structure)~\citep{radford2021learning,stark20223d}.
However, such a naive adoption fails to capture the polymorphic nature of crystallography:
\textit{A single composition can result in multiple distinct structures due to the diverse atomic arrangements, leading to significantly different physical, and chemical properties}~\citep{bernstein2020polymorphism}.
That is, the relationship between the representations $\mathbf{z}^a$ and $\mathbf{z}^b$ constitutes a one-to-many mapping rather than a one-to-one mapping, leading to inherent uncertainties in the compositional representation $\mathbf{z}^a$.

\smallskip
\noindent \textbf{Probabilistic Representation.} 
To this end, we propose to learn a probabilistic representation of composition $\mathbf{z}^a$, which naturally exhibits uncertainties of the representation, inspired by the recent Hedge Instance Embeddings (HIB)~\citep{oh2018modeling}.
The main idea here is to learn the Gaussian representation of composition, which reveals the distribution of polymorphic structures $\mathcal{P}^a$ in representation space.
Intuitively, the variance of this distribution reflects the range of diversity within these structures, thereby providing an insight into the level of uncertainty associated with composition.
More formally, we model each composition as a parameterized Gaussian distribution with learnable mean vectors and diagonal covariance matrices as follows:
\begin{equation}
    \small
    p(\Tilde{\mathbf{z}}^a|\mathbf{X}^a, \mathbf{A}^a) \sim \mathcal{N}(\mathbf{z}_{\mu}^a, \mathbf{z}_{\sigma}^a),
    \label{eq: mu_sigma} 
\end{equation}
where $\mathbf{z}_{\mu}^a = f^a_{\mu}(\mathbf{Z}^a)$ and $\mathbf{z}_{\sigma}^a = f^a_{\sigma}(\mathbf{Z}^a)$.
Here, $\mathbf{z}_{\mu}^a, \mathbf{z}_{\sigma}^a \in \mathbb{R}^{F}$ denote the mean vector and the diagonal entries of the covariance matrix, respectively, and $f^a_{\mu}$ and $f^a_{\sigma}$ refer to the modules responsible for calculating the mean and diagonal covariance matrices, respectively.
During training, we adopt the re-parameterization trick~\citep{kingma2013auto} to obtain samples from the distribution, i.e., $\Tilde{\mathbf{z}}^a = diag(\sqrt{\mathbf{z}_{\sigma}^a}) \cdot \epsilon + \mathbf{z}_{\mu}^a$, where $\epsilon \sim \mathcal{N}(0, 1)$.
While mean and variance are obtained from the shared $\mathbf{Z}^a$, we utilize different attention-based set2set pooling functions for $f^a_{\mu}$ and $f^a_{\sigma}$~\citep{vinyals2015order},
since the attentive aspects involved in calculating the mean and variance should be independent of each other.
We provide further details on the probabilistic composition encoder in Appendix~\ref{App: Probabilistic Composition Encoder}.


\subsection{Model Training via Sampled Representation Alignment}

To incorporate the structural information into the compositional representation, we define a matching probability between the composition graph $\mathcal{G}^a$ and its corresponding set of polymorphic crystal structural graphs $\mathcal{P}^{\mathcal{G}^a}$ in the Euclidean space as follows:
\begin{equation}
\small
    p(m|\mathcal{G}^a, \mathcal{P}^{\mathcal{G}^a}) \approx \sum_{p \in \mathcal{P}^{\mathcal{G}^a}}\frac{1}{J}\sum^{J}_{j=1}{\text{sigmoid}\big(-c{\|\Tilde{\mathbf{z}}_j^a - \mathbf{z}_p^b \|}_2 + d \big)},
    \label{eq: match prob}
\end{equation}
where $\Tilde{\mathbf{z}}_j^a$ is the sampled compositional representation, $\mathbf{z}_p^b$ is the structural graph representation, $c, d>0$ are parameters learned by the model for soft threshold in the Euclidean space, $J$ is the number of samples sampled from the distribution, and $\text{sigmoid}(\cdot)$ is the sigmoid function.
Moreover, $m$ is the indicator of value 1 if $\mathcal{P}^{\mathcal{G}^a}$ is the set of polymorphic structures corresponding to $\mathcal{G}^a$ and 0 otherwise.

Then, we apply the soft contrastive loss~\citep{oh2018modeling,chun2021probabilistic} as:
\begin{equation}
\small
    \mathcal{L}_{\text{con}} = 
    \begin{cases*}
    - \log{p(m|\mathcal{G}^a, \mathcal{P}^{\mathcal{G}^{a \prime}}}), & if $a = a^{\prime}$, \\
    - \log{(1 - p(m|\mathcal{G}^a, \mathcal{P}^{\mathcal{G}^{a \prime}}}), & otherwise. \\
    \end{cases*}
    \label{eq: soft con}
\end{equation}
Intuitively, the above loss aims to minimize the distance between a sampled compositional representation and its associated polymorphic structural representations, while maximizing the distance between others.
By doing so, \proposed~learns a probabilistic compositional representation that considers the structural information and its associated uncertainties, which tend to increase when multiple structures are associated with a single composition, i.e., polymorphism.

In addition to the soft contrastive loss, we utilize a KL divergence loss between the learned composition distributions and the standard normal distribution $\mathcal{N}(0, 1)$, i.e., $\mathcal{L}_{\text{KL}} = \text{KL}(p(\Tilde{\mathbf{z}}^a|\mathbf{X}^a, \mathbf{A}^a) ~ \| ~ \mathcal{N}(0,1))$, which prevents the learned variances from collapsing to zero. Therefore, our final loss for model training is given as follows: 
\begin{equation}
\small
    \mathcal{L}_{\text{total}} = \mathcal{L}_{\text{con}} + \beta \cdot \mathcal{L}_{\text{KL}},
    \vspace{-0.5ex}
    \label{eq: total}
\end{equation}
where $\beta$ is the hyperparameter for controlling the weight of the KL divergence loss.
During the inference, we use the mean vector $\mathbf{z}_{\mu}^a$ as the compositional representation and the geometric mean of diagonal covariance matrices $\mathbf{z}_{\sigma}^a$ as uncertainty~\citep{chun2021probabilistic}.

\subsection{Datasets}
\label{app:datasets}



In this section, we provide further details on the dataset used for experiments.
We first introduce the datasets utilized for the main manuscript, which is mainly based on wet-lab experiments.

\begin{itemize}[leftmargin=.1in]

\item \textbf{Materials Project}~\citep{jain2013commentary} is an openly accessible database that provides material properties calculated using density functional theory (DFT).
We have gathered 80,162 distinct compositions along with their corresponding 112,183 crystal structures computed using DFT, 
with up to 32,021 compositions having multiple potential structures.

\item \textbf{Band Gap}~\citep{zhuo2018predicting} dataset comprises experimentally determined band gap properties for non-metallic materials. 
It encompasses 2,482 distinct compositions and a total of 3,895 experimental band gap values. 
Within this dataset, 1,413 instances of duplicate experimental band gap measurements for compositions were identified. 
Consequently, our task involves predicting the band gap properties for these 2,482 compositions, with the average value being computed in cases where duplicate experimental results exist for a given composition.

\item \textbf{Formation Enthalpies}~\citep{kim2017experimental} dataset consists of experimentally determined formation enthalpy values for intermetallic phases and other inorganic compounds. 
It includes 1,141 unique compositions and a total of 1,276 experimental formation enthalpy values. 
Within this dataset, 135 cases of duplicate experimental formation enthalpy measurements for compositions were identified. 
Therefore, our objective is to predict the formation enthalpy properties for these 1,141 compositions, calculating the average value when duplicate experimental results are present for a particular composition.
We report MAE values multiplied by a factor of 10 for clear interpretation during all experiments. 

\item \textbf{Metallic}~\citep{gtt} dataset contains reduced glass transition temperature (Trg) for 584 unique metallic alloys. We report MAE values multiplied by a factor of 10 for clear interpretation during all experiments. 

\item \textbf{ESTM 300 K}~\citep{na2022public} dataset contains various properties of 368 thermoelectric materials that are measured in the temperature range of 295 K to 305 K, which is widely recognized as room temperature in chemistry.
Among the properties, we mainly target \textbf{electrical conductivity ($S/m$)}, \textbf{thermal conductivity ($W/mK$)}, and \textbf{Seebeck coefficient ($\mu V/K$)}.
Regarding electrical conductivity and thermal conductivity, we apply a logarithmic scaling to the target values because they exhibit significant skewness. 
Additionally, for the Seebeck coefficient, we use min-max scaling on the target values due to their wide range and report MAE values multiplied by a factor of 10 for clear interpretation during all experiments. 
When calculating the figure of merit ($Z\Bar{T}$) with predicted properties, 
we reverse the scaling to return the original scale and then compute it.

\item \textbf{ESTM 600 K}~\citep{na2022public} dataset contains various properties of 188 thermoelectric materials that are measured in the temperature range of 593 K to 608 K, which is widely recognized as high temperature in chemistry.
The properties we are targeting and the preprocessing steps applied are identical to those used for the \textbf{ESTM 300 K} dataset.

\end{itemize}

In addition to the wet-lab experimental datasets, we use the following seven \textbf{Matbench} datasets that contain properties from DFT calculation.

\begin{itemize} [leftmargin=.1in]
    \item \textbf{Castelli Perovskites}~\citep{castelli2012new} dataset contains formation energy of Perovskite cell of 18,928 materials.
    \item \textbf{Refractive Index}~\citep{jain2013commentary} dataset contains a refractive index of 4,764 materials, provided in \textbf{MP} database.
    \item \textbf{Shear Modulus}~\citep{jain2013commentary} dataset contains shear modulus of 10,987 materials, provided in \textbf{MP} database.
    \item \textbf{Bulk Modulus}~\citep{jain2013commentary} dataset contains bulk modulus of 10,987 materials, provided in \textbf{MP} database.
    \item \textbf{Exfoliation Energy}~\citep{choudhary2017high} dataset contains exfoliation energy 636 materials.
    \item \textbf{MP Band gap}~\citep{jain2013commentary} dataset contains band gap of 106,113 materials, provided in \textbf{MP} database.
    \item \textbf{MP Formation Energy}~\citep{jain2013commentary} dataset contains formation energy per atom in 132,752 materials, provided in \textbf{MP} database.
\end{itemize}

Following previous work~\citep{wang2021compositionally, goodall2020predicting}, we choose the target value associated with the lowest formation enthalpy for duplicate compositions found in both the MP datasets, while we use the mean of the target values for other datasets.

\subsection{Baseline Methods}
\label{sec:baseline_appendix}
In this section, we elaborate on baseline methods. 
For a fair comparison, all these baseline methods leverage the same neural network architecture and only differ in training objective function. 
\begin{itemize}[leftmargin=.1in]

\item \textbf{Rand init.} refers to the randomly initialized composition encoder without any training process. 

\item \textbf{GraphCL}~\citep{you2020graph} is a general graph-level contrastive learning strategy that uses random augmentation to construct positive and negative samples. In this paper, it learns the compositional representation based on the random augmentation on the composition graph $\mathcal{G}^a$, without utilizing structural information. For the $n$-th data in the minibatch ($N$ data points), the loss function is defined as follows follows:
\begin{equation}
l_n = -\log \frac{\exp\{\text{sim}(z_{n}, z_{n})/\tau\}}{\sum_{n'=1,n'\neq n}^{N} \exp\{\text{sim}(z_{n}, z_{n'})/\tau\} }, \\ 
\end{equation}
where $\text{sim}(\cdot, \cdot)$ indicates cosine similarity between two latent vectors. $\tau>0$ denotes temperature and is a hyperparameter. $z_i$ is the representation of the $i$-th data.

\item \textbf{MP Band~G.} and \textbf{MP Form.~E.} learn the compositional representation by predicting the DFT-calculated properties, i.e., band gap and formation energy per atom, respectively. More formally, model is trained with MAE loss for $n$-th data point in the minibatch ($N$ data points) as follows:
\begin{equation}
l_n = | Y_n - \hat{Y}_n |, \\ 
\end{equation}
where $Y_n$ and $\hat{Y}_n$ denote DFT-calculated property and model prediction, respectively.

\item \textbf{3D Infomax}~\citep{stark20223d} proposes to enhance model prediction on 3D molecular graphs by integrating 3D information of the molecules in its latent representations.
Instead of 2D molecular graphs, we learn the representation of composition graph $\mathcal{G}^a$ by maximizing the mutual information with structural graph $\mathcal{G}^b$. 
More specifically, we train the model with NTXent (Normalized Temperature-scaled Cross Entropy) loss~\citep{chen2020simple}, which is defined for $n$-th data point in minibatch of size $N$ as follows:
\begin{equation}
l_n = -\log \frac{\exp\{\text{sim}(z_{n}^a, z_{n}^b)\}}{\sum_{n' = 1, n'\neq n}^{N} \exp\{\text{sim}(z_{n}^a, z_{n'}^b)\}}, \\ 
\end{equation}
where $\text{sim}(\cdot, \cdot)$ indicates cosine similarity between two latent vectors.
\end{itemize}

Even though the primary focus of this paper is to introduce training strategies for composition encoders without any label information, we also conduct a comparative analysis of our proposed approach with previous supervised compositional representation learning methods~\citep{goodall2020predicting,wang2021compositionally}. 
Note that these works propose sophisticated model architectures for compositional representation learning, not training strategy.

\begin{itemize}[leftmargin=.1in]
    \item \textbf{Roost}~\citep{goodall2020predicting} first proposes to utilize GNNs for compositional representation learning by presenting composition as a fully connected graph, whose nodes are unique elements in composition. 
    This approach allows the model to acquire distinct and material-specific representations for each element, enabling it to capture physically meaningful properties and interactions.
    \item \textbf{CrabNet}~\citep{wang2021compositionally} designs a Transformer self-attention mechanism~\citep{vaswani2017attention} to adaptively learn the representation of individual elements based on their chemical environment. 
\end{itemize}

\subsection{Evaluation Protocol}
\label{App:evaluation protocol}

\noindent\textbf{Evaluation Metrics.}
We mainly compare the methods in terms of Mean Absolute Error (MAE) following previous work~\citep{goodall2020predicting}. 
Moreover, we provide the model performance in terms of $R^2$ in Appendix~\ref{App:additional Experiments}, which provides an intuitive measure of the fraction of the overall variance in the data that the model can account for.

During evaluation, we evaluate models in two different settings, i.e., representation learning and transfer learning. 
In both scenarios, we evaluate the model under a 5-fold cross-validation scheme, i.e., the dataset is randomly split into 5 subsets, and one of the subsets is used as the test set while the remaining subsets are used to train the model.

\noindent\textbf{Representation Learning.}
For representation learning scenarios, we fix the model parameters (i.e., $f^a$, $f_{\mu}^a$, and $f_{\sigma}^a$) and train a three-layer MLP head with LeakyReLU non-linearity to evaluate the composition obtained by various models.
Following previous works~\citep{velivckovic2018deep, you2020graph}, we train the MLP head with Adam optimizer with a fixed learning rate of 0.001 for 300 epochs.

\noindent\textbf{Transfer Learning.}
For transfer learning scenarios, we allow the model parameters (i.e., $f^a$, $f_{\mu}^a$, and $f_{\sigma}^a$) to be trained with labels in downstream tasks, jointly with a three-layer MLP head with LeakyReLU non-linearity.
During the transfer learning stage, we train the model parameters and head with the Adam optimizer for 500 epochs.
We tune the learning rate in the range of $\{0.005, 0.001, 0.0005, 0.0001\}$ with a validation set which is a subset ($20 \%$) of the training set. 
Due to the lack of data, we select the learning rate that yields the optimal performance on the validation set. 
Subsequently, we retrain the model using both the training set and the validation set, with the corresponding learning rate.

\section{Results}

\subsection{Quantitative Results}
\label{sec: Empirical Results}

\begin{table*}[t]
\caption{Representation learning performance (MAE) (Prop.: Property / Str.: Structure / Poly.: Polymorphism / Band G.: Band Gap / Form.~E.: Formation Entalphies / E.C.: Electrical Conductivity / T.C.: Thermal Conductivity).}
    \centering
    \resizebox{0.95\linewidth}{!}{
    \begin{tabular}{lccccccccccccccccc}
    \toprule
    \multirow{3}{*}{\textbf{Model}} & \multicolumn{2}{c}{\textbf{DFT}} & \multirow{3}{*}{\textbf{Poly.}} & \multirow{3}{*}{\textbf{Band G.}} & \multirow{3}{*}{\textbf{Form.~E.}} & \multirow{3}{*}{\textbf{Metallic}} & &\multicolumn{3}{c}{\textbf{ESTM 300K}} & &\multicolumn{3}{c}{\textbf{ESTM 600K}} & &\multicolumn{2}{c}{$Z\Bar{T}$} \\
    \cmidrule{2-3} \cmidrule{9-11} \cmidrule{13-15} \cmidrule{17-18}
    & \textbf{Prop.}& \textbf{Str.} & & & &  & & E.C. & T.C. & Seebeck &  & E.C. & T.C. & Seebeck &  & 300K & 600K \\
    \midrule
    \multirow{2}{*}{Rand init.} & \multirow{2}{*}{\xmark} & \multirow{2}{*}{\xmark} & \multirow{2}{*}{\xmark} &0.439  &0.671  &0.211 & & 1.029 & 0.225 & 0.451 & & 0.714  & 0.218 & 0.437 &  & 0.099 & 0.261 \\
    & & & &\scriptsize{(0.014)}  &\scriptsize{(0.066)}  &\scriptsize{(0.023)} & &\scriptsize{(0.119)}  &\scriptsize{(0.030)}  &\scriptsize{(0.031)} & & \scriptsize{(0.113)} & \scriptsize{(0.024)}  &\scriptsize{(0.087)} & &\scriptsize{(0.017)} &\scriptsize{(0.160)} \\
    \multirow{2}{*}{GraphCL} & \multirow{2}{*}{\xmark} & \multirow{2}{*}{\xmark} & \multirow{2}{*}{\xmark} &0.437  &0.677  &0.212 & &1.057 & 0.229 & 0.459 & & 0.695 & 0.206 & 0.440 &  & 0.121 & 0.211 \\
    & & & &\scriptsize{(0.022)}  &\scriptsize{(0.030)}  &\scriptsize{(0.019)} & & \scriptsize{(0.115)}  &\scriptsize{(0.040)}  &\scriptsize{(0.044)} & & \scriptsize{(0.119)}  &\scriptsize{(0.027)}  &\scriptsize{(0.077)} & &\scriptsize{(0.027)} &\scriptsize{(0.043)} \\
    \multirow{2}{*}{MP Band G.} & \multirow{2}{*}{\cmark} & \multirow{2}{*}{\xmark} & \multirow{2}{*}{\xmark} & \textbf{0.403} &0.690 &0.212 & & 1.008 &0.225  & 0.443 & & 0.690 & 0.217 & 0.436 &  & 0.129 & 0.251 \\
    & & & &\scriptsize{(0.011)}  &\scriptsize{(0.043)}  &\scriptsize{(0.028)} &  &\scriptsize{(0.081)}  &\scriptsize{(0.026)}  &\scriptsize{(0.074)} & & \scriptsize{(0.085)} &\scriptsize{(0.023)}  &\scriptsize{(0.075)} & &\scriptsize{(0.044)} &\scriptsize{(0.161)} \\
    \multirow{2}{*}{MP Form.~E.} & \multirow{2}{*}{\cmark} & \multirow{2}{*}{\xmark} & \multirow{2}{*}{\xmark} & 0.416 & 0.619 & 0.203 & & 1.121 & 0.228 & 0.441 & & 0.784 & 0.220 & 0.444 &  & 0.093 & 0.328 \\
    & & & &\scriptsize{(0.017)}  &\scriptsize{(0.062)}  &\scriptsize{(0.022)} & & \scriptsize{(0.137)} &\scriptsize{(0.024)}  &\scriptsize{(0.078)} & & \scriptsize{(0.078)}& \scriptsize{(0.021)} & \scriptsize{(0.091)} & & \scriptsize{(0.008)} & \scriptsize{(0.075)} \\
    \multirow{2}{*}{3D Infomax} & \multirow{2}{*}{\xmark} & \multirow{2}{*}{\cmark} & \multirow{2}{*}{\xmark} & 0.428  & 0.654 &0.201 & &0.969  & 0.217 &0.432  & & 0.692 & 0.212 & 0.428 &  & 0.105 & 0.171 \\
    & & & &\scriptsize{(0.015)}  &\scriptsize{(0.032)}  &\scriptsize{(0.032)}  & &\scriptsize{(0.110)}  &\scriptsize{(0.040)}  &\scriptsize{(0.070)} & &\scriptsize{(0.102)} &\scriptsize{(0.013)}  &\scriptsize{(0.076)} & &\scriptsize{(0.030)} &\scriptsize{(0.023)} \\
    \midrule
    \multirow{2}{*}{\proposed} & \multirow{2}{*}{\xmark} & \multirow{2}{*}{\cmark} & \multirow{2}{*}{\cmark} &0.407  & \textbf{0.592}  & \textbf{0.194} & & \textbf{0.912} & \textbf{0.197}  & \textbf{0.388} & & \textbf{0.665} & \textbf{0.189} & \textbf{0.412} &  & \textbf{0.070} & \textbf{0.168}\\
    & & & &\scriptsize{(0.013)}  &\scriptsize{(0.039)}  &\scriptsize{(0.017)} &  &\scriptsize{(0.121)}  &\scriptsize{(0.020)}  &\scriptsize{(0.059)} & &\scriptsize{(0.126)}  &\scriptsize{(0.017)}  &\scriptsize{(0.043)} & & \scriptsize{(0.014)} & \scriptsize{(0.021)}\\
    \bottomrule
    \end{tabular}}
    \label{tab: main table}
\end{table*}

\noindent\textbf{Representation Learning.}
In Table~\ref{tab: main table}, we have the following observations:
\textbf{1)} We find out that utilizing structural information (\textbf{\small{Str.}} \cmark) generally learns more high-quality compositional representations compared with those that overlook the structural information (\textbf{\small{Str.}} \xmark).
This is consistent with the established knowledge in crystallography, which emphasizes that structural details, including crystal structure and symmetry, play a crucial role in determining a wide range of physical, chemical, and mechanical properties~\citep{bernstein2020polymorphism,braga2009crystal}.
\textbf{2)} Moreover, we observe \proposed~outperforms baseline methods that overlook polymorphism in their model design (\textbf{\small{Poly.}} \xmark).
This highlights the significance of our probabilistic approach, which not only offers insights into polymorphism-related uncertainties but also yields high-quality representations.
\textbf{3)} On the other hand, we notice that utilizing DFT-calculated values contributes to the model's understanding of a specific target property (see \textbf{\small{Prop.}} \cmark). 
For instance, when the model is trained with a DFT-calculated band gap (i.e., MP Band G.), it surpasses all other models when predicting specifically the experimental band gap values. 
{This highlights that knowledge acquired from DFT-calculated properties can be helpful in predicting the properties obtained from wet-lab experiments.}
However, these representations are highly tailored to a particular target property, which restricts their generalizability for diverse tasks.
We provide visualization of representation and statistical significance of improvement in Appendix~\ref{App: Representation Space Analysis} and~\ref{App: Statistical Significance Test}, respectively.

\smallskip
\noindent \textbf{Physical Validity of Predicted Properties.}
To further verify the physical validity of predicted properties, we theoretically calculate the figure of merit $Z\Bar{T}$
\footnote{
In thermoelectric materials, the figure of merit $Z\Bar{T}$ plays a fundamental role in determining how effectively power can be generated and energy can be harvested across various applications~\citep{nozariasbmarz2020review}.
}
of thermoelectrical materials with the model-predicted properties in ESTM datasets in Table~\ref{tab: main table}.
Note that the model-predicted properties are assumed to be physically valid, if the theoretically calculated figure of merit $Z\Bar{T}$ is accurate (i.e., low MAE for $Z\Bar{T}$ in Table \ref{tab: main table}).
More specifically, given predicted electrical conductivity (E.C.) $\sigma$, thermal conductivity (T.C.) $\lambda$, Seebeck coefficient $S$, we can compute the figure of merit $Z\Bar{T}$ as follows: $Z\Bar{T} = \frac{S^2 \sigma}{\lambda} \Bar{T}$, where $\Bar{T}$ indicates a conditioned temperature, i.e., 300 K and 600 K.
In Table~\ref{tab: main table}, we have following observations:
{\textbf{1)} Looking at the general model performance on ESTM datasets and $Z\Bar{T}$, we find that performing well on ESTM datasets does not necessarily indicate the predictions are physically valid. 
For instance, while training the model on DFT-calculated bandgap (MP Band G.) consistently yields more accurate predictions on the ESTM dataset compared to GraphCL, the $Z\Bar{T}$ values derived from MP Band G. predictions are significantly worse than those obtained from GraphCL.}
\textbf{2)} Overall, models that incorporate structural information tend to produce physically valid predictions in both ESTM datasets, underscoring the importance of the crystal structural information.
\textbf{3)} Moreover, \proposed~consistently outperforms baseline methods, demonstrating that \proposed~not only learns accurate representations of composition but also ensures the physical validity of the predictions.

\begin{table*}
\begin{minipage}{0.55\linewidth}{
\centering
    \includegraphics[width=0.9\linewidth]{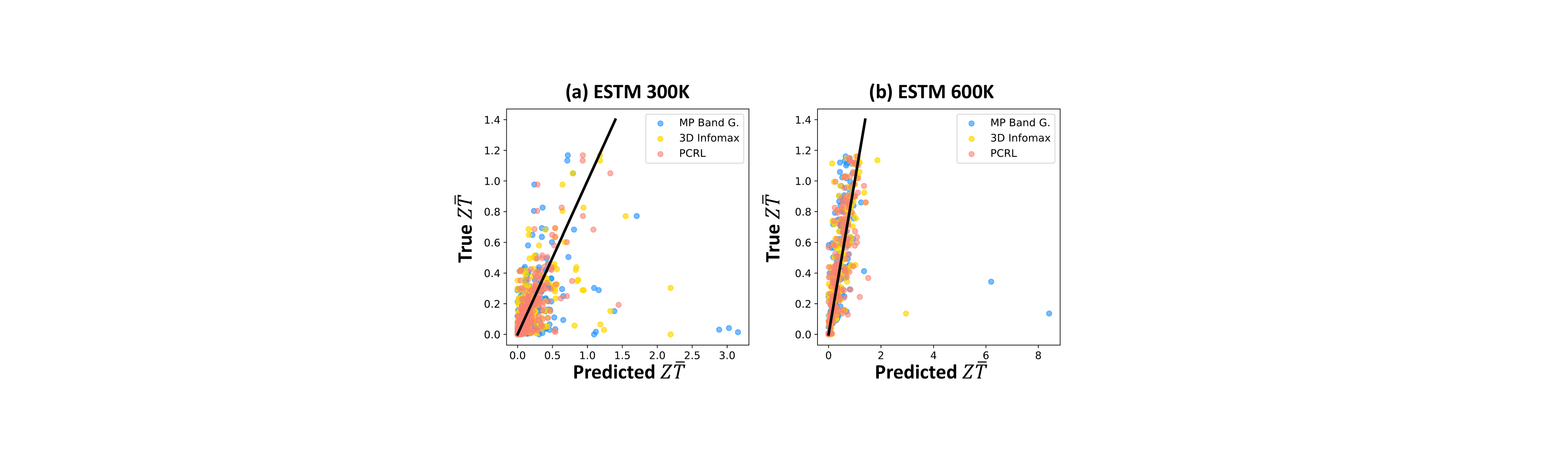}
    \vspace{4ex}
    \captionof{figure}{Scatter plot between true and predicted $Z\Bar{T}$.}
    \label{fig: scatter plot}
}\end{minipage}
\hspace{1ex}
\begin{minipage}{0.45\linewidth}{
\centering
    \includegraphics[width=0.9\linewidth]{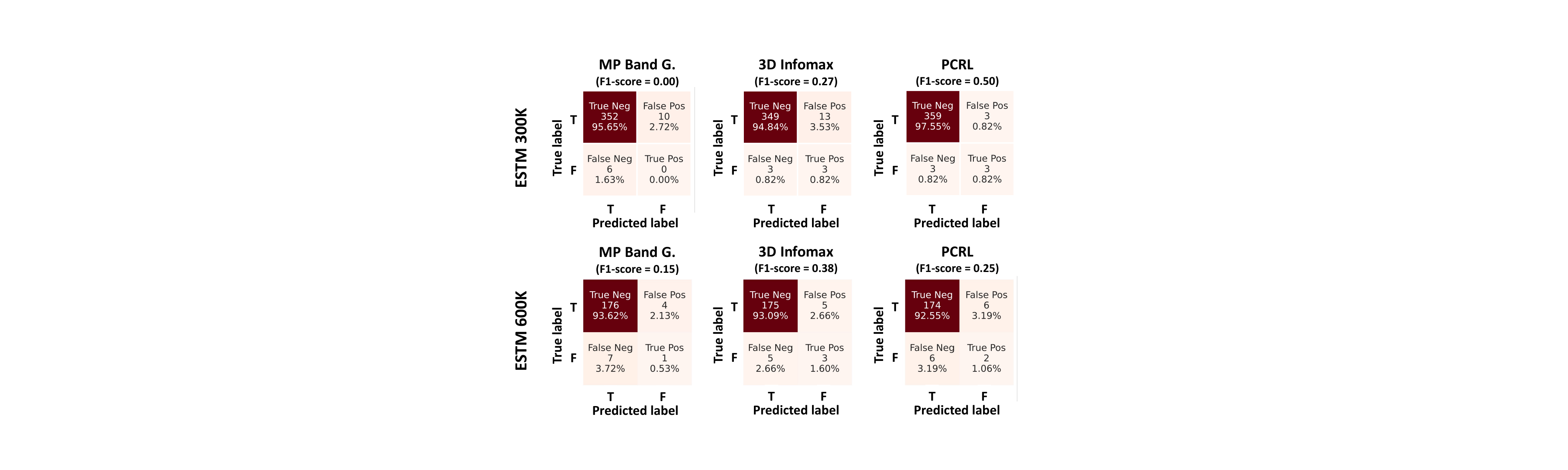}
    \captionof{figure}{High-throughput screening results.}
    \label{fig: screening}
}\end{minipage}
\end{table*}

\smallskip
\noindent \textbf{Further Analysis.}
In this section, we delve deeper into the physical validity of predicted properties for thermoelectrical materials by observing scatter plots that compare the actual ground truth values of $Z\Bar{T}$ with the values obtained by the model predictions.
For clearer visualization, we select one baseline model from models that consider DFT-calculated properties (i.e., {MP Band G.}) and structures (i.e., {3D Infomax}).
In Figure~\ref{fig: scatter plot}, we notice that the predictions produced by \proposed~consistently yield accurate calculations of $Z\Bar{T}$ without any outliers. 
This observation underscores the model's ability to predict physically valid properties for thermoelectrical materials. 
Additionally, we observe that the model, specifically {MP Band G.}, which lacks consideration of the structural information within composition, tends to produce outliers more frequently when contrasted with models that incorporate structural information.
More specifically, three outliers made by {MP Band G.} in Figure~\ref{fig: scatter plot} (a) are $\text{Co}_{9}\text{S}_{8}$, $\text{Cu}_{5}\text{Sn}_{2}\text{S}_{6.65}\text{Cl}_{0.35}$, and $\text{Cu}_{5.133}\text{Sn}_{1.866}\text{S}_{6.65}\text{Cl}_{0.35}$.
In case of $\text{Co}_{9}\text{S}_{8}$, there exist only one possible structure in {MP} dataset, and there was no existing structure for $\text{Cu}_{5}\text{Sn}_{2}\text{S}_{6.65}\text{Cl}_{0.35}$, and $\text{Cu}_{5.133}\text{Sn}_{1.866}\text{S}_{6.65}\text{Cl}_{0.35}$.
This suggests that {MP Band G.} encounters difficulty in acquiring accurate physical properties for materials where obtaining structural information is computationally challenging.
On the other hand, in Figure~\ref{fig: scatter plot} (b), two outliers made by {MP Band G.} are $\text{GeTe}$ and $\text{SnTe}$, each of which has three possible structures in {MP} dataset.
This indicates that {MP Band G.} suffers from obtaining valid physical properties from polymorphic structures.
In conclusion, we argue that this finding underscores the significance of incorporating structural information for accurate predictions.

\smallskip
\noindent \textbf{High-Throughput Screening.}
As described in the main manuscript, the figure of merit $Z\Bar{T}$ determines how effectively power can be generated and energy can be harvested across various real-world applications.
To discover novel materials of high $Z\Bar{T}$, we perform high-throughput screening based on the predicted $Z\Bar{T}$ in Figure~\ref{fig: screening}.
In particular, for thermoelectrical materials at room temperature (300 K), we establish a threshold of $Z\Bar{T} = 0.8$, and for high-temperature scenarios (600 K), we use a threshold of $Z\Bar{T} = 1.1$.
We observe that \proposed~outperforms all other baseline methods in {ESTM 300K} datasets while performing competitively with {3D Infomax} in {ESTM 600K}.
This again demonstrates the importance of structural information in compositional representation learning, which has been overlooked in previous works~\citep{goodall2020predicting, wang2021compositionally}.

\begin{table*}
\caption{Representation learning performance on DFT-calculated datasets (MAE). }
    \centering
    \resizebox{0.9\linewidth}{!}{
    \begin{tabular}{lccccccccccc}
    \toprule
    \multirow{3}{*}{\textbf{Model}} & \multicolumn{2}{c}{\textbf{DFT}} & \multirow{3}{*}{\textbf{Poly.}} & \textbf{Castelli} & \textbf{Refractive} & \textbf{Shear} & \textbf{Bulk} & \textbf{Exfoliation} & & \multicolumn{2}{c}{\textbf{MP}}\\
    \cmidrule{2-3} \cmidrule{11-12}
    & \textbf{Prop.}& \textbf{Str.} & & \textbf{Perovskites} & \textbf{Index} & \textbf{Modulus} & \textbf{Modulus} & \textbf{Energy} & & \textbf{Band G.} & \textbf{Form.~E.}\\
    \midrule
    \multirow{2}{*}{Rand init.} & \multirow{2}{*}{\xmark} & \multirow{2}{*}{\xmark} & \multirow{2}{*}{\xmark} & 0.140 & 0.394 & 0.115 & 0.085 & 54.49 &  & 0.354 & 0.119 \\
    & & & &\scriptsize{(0.004)}  &\scriptsize{(0.091)}  &\scriptsize{(0.003)} & \scriptsize{(0.003)} & \scriptsize{(16.08)} &  &\scriptsize{(0.005)} & \scriptsize{(0.002)}\\
    \multirow{2}{*}{GraphCL} & \multirow{2}{*}{\xmark} & \multirow{2}{*}{\xmark} & \multirow{2}{*}{\xmark} & 0.145 & 0.386 & 0.117 & 0.084 & 56.20 &  & 0.351 & 0.121 \\
    & & & &\scriptsize{(0.006)}  &\scriptsize{(0.094)}  &\scriptsize{(0.002)} & \scriptsize{(0.002)} & \scriptsize{(17.51)} &  &\scriptsize{(0.004)} & \scriptsize{(0.001)}\\
    \multirow{2}{*}{MP Band G.} & \multirow{2}{*}{\cmark} & \multirow{2}{*}{\xmark} & \multirow{2}{*}{\xmark} & 0.141 & 0.399 & 0.116 & 0.085 & 56.14 &  & 0.354 & 0.119 \\
    & & & &\scriptsize{(0.004)}  &\scriptsize{(0.085)}  &\scriptsize{(0.002)} & \scriptsize{(0.004)} & \scriptsize{(16.86)} &  &\scriptsize{(0.007)} & \scriptsize{(0.002)}\\
    \multirow{2}{*}{MP Form.~E.} & \multirow{2}{*}{\cmark} & \multirow{2}{*}{\xmark} & \multirow{2}{*}{\xmark} & 0.134 & \textbf{0.379} & 0.108 & \textbf{0.080} & 55.40 &  & 0.338 & 0.115 \\
    & & & &\scriptsize{(0.004)}  &\scriptsize{(0.093)}  &\scriptsize{(0.002)} & \scriptsize{(0.003)} & \scriptsize{(17.29)} &  &\scriptsize{(0.002)} & \scriptsize{(0.001)}\\
    \multirow{2}{*}{3D Infomax} & \multirow{2}{*}{\xmark} & \multirow{2}{*}{\cmark} & \multirow{2}{*}{\xmark} & 0.147 & 0.388 & 0.117 & 0.088 & 54.47 &  & 0.354 & 0.116 \\
    & & & &\scriptsize{(0.004)}  &\scriptsize{(0.094)}  &\scriptsize{(0.003)} & \scriptsize{(0.004)} & \scriptsize{(15.24)} &  &\scriptsize{(0.005)} & \scriptsize{(0.002)}\\
    \midrule
    \multirow{2}{*}{\proposed} & \multirow{2}{*}{\xmark} & \multirow{2}{*}{\cmark} & \multirow{2}{*}{\cmark} & \textbf{0.132} & 0.394 & \textbf{0.107} & 0.083 & \textbf{52.90} &  & \textbf{0.328} & \textbf{0.112} \\
    & & & &\scriptsize{(0.007)}  &\scriptsize{(0.092)}  &\scriptsize{(0.002)} & \scriptsize{(0.003)} & \scriptsize{(14.00)} &  &\scriptsize{(0.005)} & \scriptsize{(0.003)}\\
    \bottomrule
    \end{tabular}}
    \label{tab: app dft-calculated table}
\end{table*}

\smallskip
\noindent \textbf{Model Performance in DFT-calculated properties.}
Although DFT-calculated properties frequently differ from actual wet-lab experimental properties~\citep{jha2019enhancing}, we have included experimental outcomes for seven DFT-calculated properties from the Matbench dataset~\citep{dunn2020benchmarking}. 
These Matbench datasets were assessed using a five-fold cross-validation approach with train/validation/test splits set at a ratio of 72/8/20, as given in previous work~\citep{wang2021compositionally}.
In Table~\ref{tab: app dft-calculated table}, we have following observations:
\textbf{1)} In the DFT-based dataset, we observed significant disparities in trends compared to the experimental datasets in Table~\ref{tab: main table}, demonstrating the inherent difference between the experimental data and DFT-calculated data.
For instance, we noticed that the {MP Form.~E.} model consistently outperforms the {MP Band G.} and {3D Infomax} models. 
\textbf{2)} Furthermore, given that the datasets are designed to pick the target value linked to the lowest formation enthalpy among different polymorphic structures for a single composition, 
we find that models trained with specific DFT-calculated values (i.e., \textbf{\small Prop. \cmark}) do not outperform models trained on corresponding datasets. 
This discrepancy is attributed to properties derived from non-lowest formation enthalpy polymorphic structures, which can introduce confusion to the model.
\textbf{3)} However, we observe \proposed~generally outperforms baseline models, demonstrating its effectiveness in not only wet-lab experimental datasets but also in DFT-calculated datasets.

\smallskip
\noindent \textbf{Fine-Tuning.}
In this section, we compare the models' performance in fine-tuning scenarios in Table~\ref{tab:transfer learning}.
{It is worth noting that the transfer of knowledge from the pre-training phase to downstream tasks, which have limited labeled data, via fine-tuning is vital in materials science due to the common data scarcity problem in wet-lab experimental data.}
We have the following observations:
{\textbf{1)} Although recently proposed supervised learning approaches (i.e., Roost and CrabNet) are specifically tailored for material property prediction solely based on compositional data with a large number of parameters, we observe randomly initialized variant of \proposed~(i.e., Rand init.) outperforms the methods.
This is because the previous works require an extensive dataset computed via DFT for model training, limiting their applicability in wet-lab experimental data, which is more realistic in real-world materials discovery.
{\textbf{2)} From the comparisons between the randomly initialized variant of~\proposed~(i.e., Rand init.) and the baseline methods in Table~\ref{tab:transfer learning}, we observe that the pre-training step sometimes detrimentally affects model performance, i.e., negative transfer.}
This indicates that without an elaborate design of the tasks, pre-training may incur negative knowledge transfer to the downstream tasks~\citep{zhang2022survey}.
\textbf{3)} {However, by comparing Rand init. with \proposed~in Table~\ref{tab:transfer learning}, we observe that pre-training the model with \proposed~objective consistently leads to performance gain in various downstream tasks, i.e., positive transfer.}
We attribute this to the probabilistic representation, which maintains a high variance for uncertain materials, thereby preventing the representations of the materials from overfitting to the pretraining task.
{\textit{Given the widespread issue of data scarcity in wet-lab experimental data, this highlights the practicality of PCRL within the materials discovery process}.}
In Appendix \ref{App: Fine-tuning OOD}, we assess the transferability of various pre-training strategies in an out-of-distribution context, where the datasets are divided according to the types of elements involved.

\begin{table*}
\begin{minipage}{0.7\linewidth}{
\centering
\caption{Fine-tuning performance (measured by MAE).}
\resizebox{0.99\linewidth}{!}{
    \begin{tabular}{lccccccccccccccccc}
    \toprule
    \multirow{3}{*}{\textbf{Model}} & \multirow{3}{*}{\textbf{Band G.}} & \multirow{3}{*}{\textbf{Form.~E.}} & \multirow{3}{*}{\textbf{Metallic}} & &\multicolumn{3}{c}{\textbf{ESTM 300 K}} & &\multicolumn{3}{c}{\textbf{ESTM 600 K}} & &\multicolumn{2}{c}{$Z\Bar{T}$} \\
    \cmidrule{6-8} \cmidrule{10-12} \cmidrule{14-15}
    & & & & & E.C. & T.C. & Seebeck &  & E.C. & T.C. & Seebeck &  & 300K & 600K \\
    \midrule
    \multicolumn{15}{l}{\textbf{Pre-training} \xmark} \\
    \midrule
    \multirow{2}{*}{Rand init.} & 0.390 & 0.599 & 0.204 & & 0.849 & 0.202 & 0.425 & & 0.659 & 0.209 & 0.402 &  & 0.073 & 0.218 \\
    &\scriptsize{(0.012)}  &\scriptsize{(0.053)}  &\scriptsize{(0.014)} & & \scriptsize{(0.174)}  &\scriptsize{(0.027)}  &\scriptsize{(0.048)} & & \scriptsize{(0.098)}  &\scriptsize{(0.019)}  &\scriptsize{(0.082)} & &\scriptsize{(0.024)} &\scriptsize{(0.076)}\\
    \multirow{2}{*}{Roost} & 0.384 & 0.743 & 0.199 & & 0.851 & 0.216 & 0.406  & & 0.684 & 0.240  & 0.402 & & 0.066 & 0.297 \\
    &\scriptsize{(0.008)}  &\scriptsize{(0.069)}  &\scriptsize{(0.023)} & & \scriptsize{(0.126)}  &\scriptsize{(0.037)}  &\scriptsize{(0.046)} & & \scriptsize{(0.180)}  &\scriptsize{(0.048)}  &\scriptsize{(0.054)} & & \scriptsize{(0.005)} & \scriptsize{(0.031)} \\
    \multirow{2}{*}{CrabNet} & 0.403  & 0.759 & 0.220   &  & 1.016  & 0.285 & 0.491 & & 0.816  & 0.309  & 0.691 & & 0.140 & 0.574  \\
    &\scriptsize{(0.008)}  &\scriptsize{(0.052)}  &\scriptsize{(0.017)} & & \scriptsize{(0.153)}  &\scriptsize{(0.049)}  &\scriptsize{(0.088)} & & \scriptsize{(0.167)}  &\scriptsize{(0.023)}  &\scriptsize{(0.057)} & & \scriptsize{(0.063)} & \scriptsize{(0.347)} \\
    \midrule
    \multicolumn{15}{l}{\textbf{Pre-training} \cmark} \\
    \midrule
    \multirow{2}{*}{GraphCL} &0.391  &0.607  &0.193 & &0.862 & 0.198 & 0.412 & & 0.643 & 0.205 & 0.412 &  & 0.064 & 0.202 \\
    &\scriptsize{(0.011)}  &\scriptsize{(0.026)}  &\scriptsize{(0.018)} & & \scriptsize{(0.236)}  &\scriptsize{(0.031)}  &\scriptsize{(0.006)} & & \scriptsize{(0.098)}  &\scriptsize{(0.021)}  &\scriptsize{(0.077)} & &\scriptsize{(0.014)} &\scriptsize{(0.086)}\\
    \multirow{2}{*}{MP Band G.} & \textbf{0.382} &0.604 &0.193 & & 0.829 &0.210  & 0.405 & & 0.632 & 0.197 & 0.402 & & 0.067 & 0.206 \\
    &\scriptsize{(0.012)}  &\scriptsize{(0.036)}  &\scriptsize{(0.025)} &  &\scriptsize{(0.187)}  &\scriptsize{(0.038)}  &\scriptsize{(0.006)} & & \scriptsize{(0.095)} &\scriptsize{(0.028)}  &\scriptsize{(0.081)} & &\scriptsize{(0.011)} &\scriptsize{(0.119)} \\
    \multirow{2}{*}{MP Form.~E.} & 0.391 & 0.582 & 0.197 & & 0.822 & 0.195 & 0.410 & & 0.641 & 0.209 & 0.428 & & 0.069 & 0.223 \\
    &\scriptsize{(0.013)}  &\scriptsize{(0.015)}  &\scriptsize{(0.019)} & & \scriptsize{(0.167)} &\scriptsize{(0.031)}  &\scriptsize{(0.041)} & & \scriptsize{(0.102)}& \scriptsize{(0.043)} & \scriptsize{(0.086)} & &\scriptsize{(0.010)} &\scriptsize{(0.097)}\\
    \multirow{2}{*}{3D Infomax} & 0.391  & 0.606 &0.194 & &0.844  & 0.210 &0.402  & & 0.633 & 0.207 & 0.391 & & 0.066 & \textbf{0.154} \\
    &\scriptsize{(0.006)}  &\scriptsize{(0.027)}  &\scriptsize{(0.019)}  & &\scriptsize{(0.195)}  &\scriptsize{(0.032)}  &\scriptsize{(0.005)} & &\scriptsize{(0.133)} &\scriptsize{(0.018)}  &\scriptsize{(0.077)}& &\scriptsize{(0.010)} &\scriptsize{(0.032)} \\
    \midrule
    \multirow{2}{*}{\proposed} &0.386  & \textbf{0.576}  & \textbf{0.191} & & \textbf{0.822} & \textbf{0.189}  & \textbf{0.386} & & \textbf{0.626} & \textbf{0.195} & \textbf{0.390} & & \textbf{0.058} & 0.179 \\
    &\scriptsize{(0.021)}  &\scriptsize{(0.042)}  &\scriptsize{(0.024)} &  &\scriptsize{(0.162)}  &\scriptsize{(0.037)}  &\scriptsize{(0.069)} & &\scriptsize{(0.161)}  &\scriptsize{(0.015)}  &\scriptsize{(0.077)} & &\scriptsize{(0.012)}  &\scriptsize{(0.052)} \\
    \bottomrule
    \end{tabular}}
    \label{tab:transfer learning}
}\end{minipage}
\hspace{0.3ex}
\begin{minipage}{0.27\linewidth}{
\centering
    \includegraphics[width=0.9\linewidth]{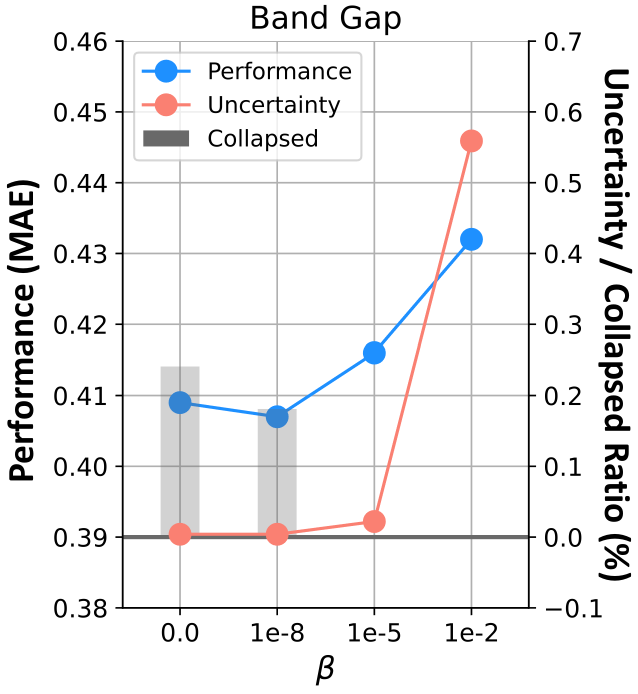}
    \captionof{figure}{Sensitivity Analysis on $\beta$.}
    \label{fig:sensitivity}
}\end{minipage}
\end{table*}

\smallskip
\noindent \textbf{Effect of the Hyperparameter $\beta$.}
In this section, we verify the empirical effect of the hyperparameter $\beta$, which controls the weight of the KL divergence loss computed between the learned distributions and the standard normal distribution, in Equation~\ref{eq: total}.
We have the following observations from Figure~\ref{fig:sensitivity}: 
\textbf{1)} As the hyperparameter $\beta$ increases, the average variance of the learned distributions (i.e., uncertainty) also increases,
and the dimension of the variance vectors that collapse to zero (i.e., collapsed ratio) decreases. This indicates that the KL divergence loss effectively prevents the distributions from collapsing. 
\textbf{2)} On the other hand, the performance of~\proposed~deteriorates as $\beta$ increases, indicating that emphasizing the KL divergence loss too much causes \proposed~to struggle in learning high-quality compositional representations. 
However, reducing $\beta$ does not always result in improved performance, as collapsed distribution may not effectively capture information from polymorphic structures.
Hence, selecting an appropriate value of $\beta$ is vital for learning high-quality compositional representations while maintaining a suitable level of uncertainty.
This selection process could be a potential limitation, as it requires a trial-and-error approach to determine the optimal value.
We provide results of additional ablation studies and analysis on hyperparameters in Appendix~\ref{App: Model Analysis}.


\begin{wrapfigure}{rt}{0.35\textwidth}
\captionof{table}{Representation learning performance (MAE) comparison trained with different datasets.}
    \resizebox{0.95\linewidth}{!}{
    \raisebox{0pt}[\dimexpr\height-0.5\baselineskip\relax]{
    \begin{tabular}{lccc}
    \toprule
    \textbf{Dataset} & \textbf{Band G.} & \textbf{Form. E.} & \textbf{Metallic} \\
    \midrule
    MP & \textbf{0.407}  & \textbf{0.592}  & \textbf{0.194} \\
    (Ours)& \scriptsize{(0.013)}  &\scriptsize{(0.039)}  &\scriptsize{(0.017)} \\
    \midrule
    \multirow{2}{*}{OQMD} & 0.425  & 0.663  & \textbf{0.194} \\
    & \scriptsize{(0.011)}  &\scriptsize{(0.035)}  &\scriptsize{(0.028)} \\
    \bottomrule
    \end{tabular}}}
    \label{tab: different database}
\end{wrapfigure}

\smallskip
\noindent\textbf{Pre-training with Other Databases.} While our model is initially pre-trained using the MP database, we also explore whether other databases can enhance its performance.
To do so, we pre-train \proposed~with another database, i.e., Open Quantum Materials Database (OQMD)\footnote{\label{url: oqmd}\url{https://oqmd.org/}}, which consists of 640,964 unique compositions and their corresponding 1,222,097 structures.
It is widely known that the MP database primarily comprises crystal structures that exist in the real world, whereas the OQMD database includes both existing crystal structures and theoretically possible structures.
In Table \ref{tab: different database}, we observe that training with the OQMD database deteriorates model performance, suggesting that training with theoretically possible structures may confuse the model when applied to wet-lab experimental datasets that comprise real-world materials. 
Consequently, we advocate for pre-training the model with the MP database as a beneficial strategy for the real-world material discovery process.
We provide full experimental results on various downstream datasets in Appendix \ref{App: Pre-training on OQMD Dataset}.

\begin{wrapfigure}{rt}{0.35\textwidth}
\vspace{-2ex}
\captionof{table}{Representation learning performance (MAE) comparison trained with different relationship.}
    \resizebox{0.95\linewidth}{!}{
    \begin{tabular}{lccc}
    \toprule
    \textbf{Relationship} & \textbf{Band G.} & \textbf{Form. E.} & \textbf{Metallic} \\
    \midrule
    One-to-Many & \textbf{0.407}  & \textbf{0.592}  & \textbf{0.194} \\
    (Ours) & \scriptsize{(0.013)}  &\scriptsize{(0.039)}  &\scriptsize{(0.017)} \\
    \midrule
    \multirow{2}{*}{Many-to-Many} & 0.410  & 0.617  & 0.210 \\
    & \scriptsize{(0.012)}  &\scriptsize{(0.025)}  &\scriptsize{(0.023)} \\
    \bottomrule
    \end{tabular}}
    \label{tab: different relationship}
\end{wrapfigure}

\smallskip
\noindent \textbf{Many-to-many Relationship between Composition and Structures.}
In material science, polymorphism refers to the phenomenon where a single composition can be related to multiple crystal structures.
On the other hand, a single structure can also be related to multiple compositions; different materials may share the same atomic arrangement, known as isostructural \cite{coles2014same}, resulting in a many-to-many relationship between composition and structure.
To this end, we examine whether modeling many-to-many relationships would be beneficial by learning the probabilistic representation of crystal structure.
In Table \ref{tab: different relationship}, we observe that modeling many-to-many relationships between the composition and structure deteriorates model performance.
We attribute this to the wet-lab experimental environment, where uncertainties related to isostructural materials are absent due to comprehensive knowledge of the material's composition. 
Consequently, we contend that addressing the ambiguities arising from polymorphism, rather than isostructural issues, is more vital for real-world material discovery.
We also provide full experimental results on various downstream datasets in Appendix \ref{App: Many-to-many relationship}.

\subsection{Uncertainty Analysis}
\label{sec: Uncertainty Analysis}


\begin{table*}
\begin{minipage}{0.5\linewidth}{
\centering
    \includegraphics[width=0.99\linewidth]{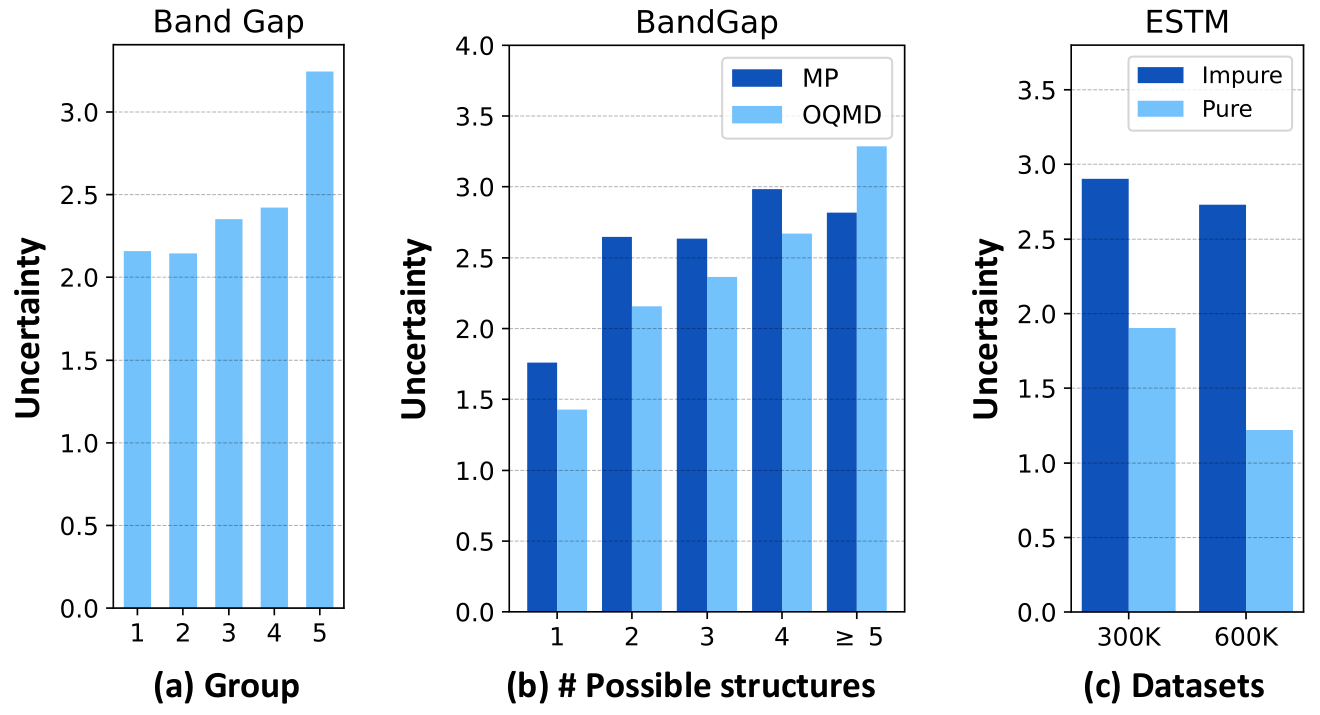}
    \captionof{figure}{Uncertainty analysis.}
    \label{fig:uncertainty}
}\end{minipage}
\hspace{0.3ex}
\begin{minipage}{0.5\linewidth}{
\centering
    {\includegraphics[width=0.9\linewidth]{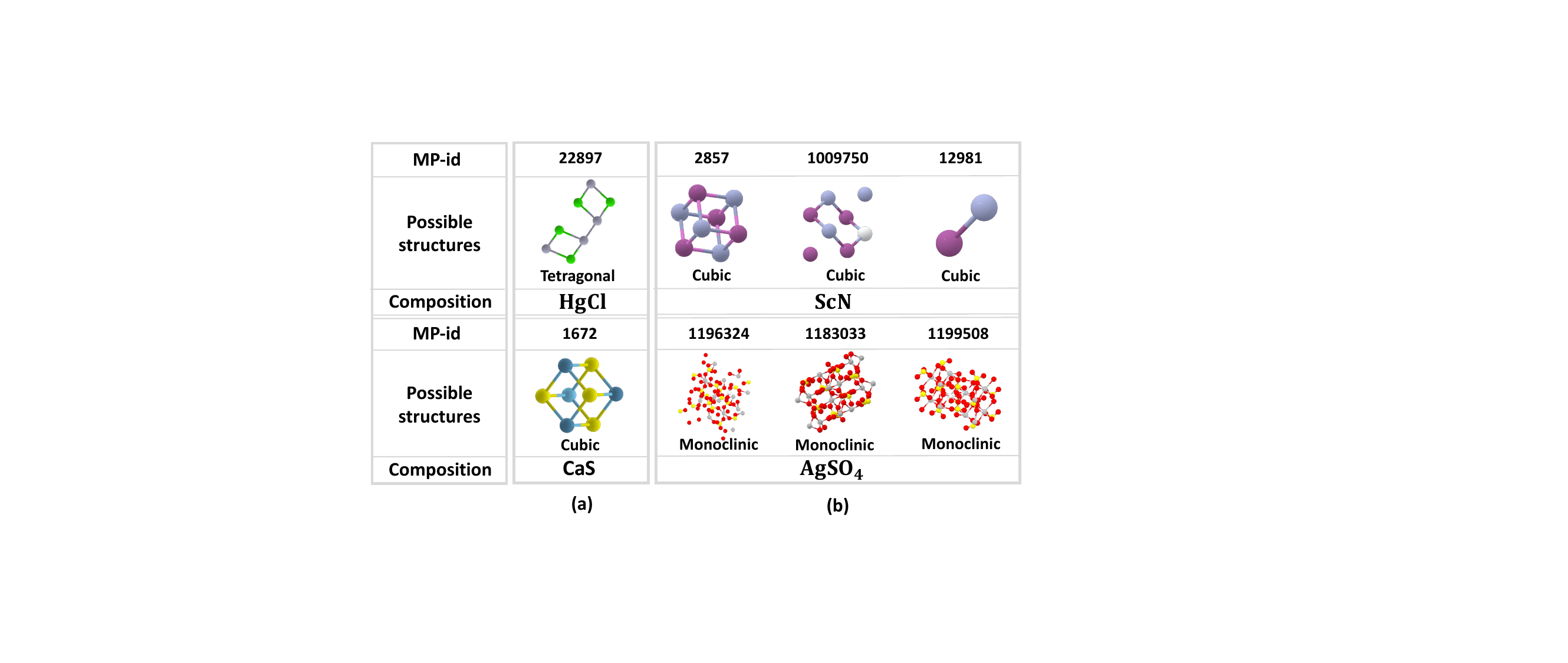}}
    \caption{Qualitative uncertainty analysis.}
    \label{fig:case study}
}\end{minipage}
\vspace{-2ex}
\end{table*}

\textbf{Model Accuracy.}
First, we analyze the correlation between the accuracy of model predictions and the uncertainties present in the Band Gap dataset.
To achieve this, we initially categorize composition based on MAE into intervals such as 0.0 to 1.0, 1.0 to 2.0, $\cdots$, and 4.0 to 5.0.
For example, Group 1 in Figure~\ref{fig:uncertainty} (a) contains the set of compositions whose MAEs are in the range from 0.0 to 1.0.
We then calculate the average uncertainties of the model for each group.
In Figure~\ref{fig:uncertainty} (a), uncertainty increases as the MAE values increase, 
demonstrating that the model effectively estimates uncertainties associated with the prediction.

\smallskip
\noindent \textbf{Number of Potential Structures.}
Moreover, we examine how uncertainties vary according to the number of potential structures. 
To do so, we first collect all possible structures of composition in Band Gap dataset from MP database \footnote{\label{url: mp}\url{https://next-gen.materialsproject.org/}}, and OQMD \textsuperscript{\ref{url: oqmd}}.
Subsequently, we compute the average uncertainties for compositional groups with the same number of possible structures.
In Figure~\ref{fig:uncertainty} (b), we have the following observations:
\textbf{1)} In general, the uncertainty of a composition that has polymorphic structures ($\#$ possible structures $\geq 2$) is higher than that of the composition with a single structure ($\#$ possible structures $= 1$), demonstrating that~\proposed~learns uncertainties regarding polymorphic structures.
\textbf{2)} Moreover, an increase in the number of possible structures generally leads to an increase in the uncertainty, demonstrating that \proposed~learns uncertainties related to the diverse polymorphic structures.
Note that this trend is mainly shown in the OQMD dataset, encompassing not only realistic but also theoretically possible structures, indicating that \proposed~acquires knowledge of theoretical uncertainties in materials science.
We provide case studies demonstrating that \proposed~acquires detailed insights concerning polymorphic structures in Section~\ref{sec: Qualitative Analysis}.

\smallskip
\noindent \textbf{Impurity of Materials.}
Next, we investigate how impurities in materials influence the uncertainty in composition.
Specifically, we compare the average composition uncertainty between groups of doped or alloyed materials (i.e., Impure) and their counterparts (i.e., Pure) in thermoelectric materials datasets, i.e., ESTM 300K and ESTM 600K, where doping and alloying are commonly employed to enhance the {thermoelectric performance of materials.}
In Figure~\ref{fig:uncertainty} (c), we notice a substantial increase in the uncertainty within impure materials compared with their pure counterparts. 
This observation is in line with common knowledge in materials science that doping or alloying can lead to chaotic transformations in a stable structure~\citep{kawai1992effects,jin2014doping}, demonstrating that \proposed~also captures the complexity of structure as the uncertainty.

\smallskip
\noindent \textbf{Materials Discovery with Uncertainties.}
In materials science, the presence of many possible structures for a given composition indicates that researchers have a higher degree of freedom to manipulate the substance, allowing for a wider range of applications for the material.
By leveraging the uncertainties identified by \proposed, it becomes possible to ascertain if a given composition might present multiple potential structures, thus opening up a broad spectrum of applications without relying on expensive DFT calculations.
In this section, we study the materials for which our model provides high uncertainty, specifically focusing on those with five polymorphic structures.
\textbf{Cesium Iodide (CsI)} is a material that has the highest uncertainty among the materials that have five polymorphic structures.
Notably, CsI is known for its high scintillation efficiency, meaning that it can effectively convert high-energy radiation (e.g., X-rays or gamma rays) into visible light \citep{zhao2004x}. 
This characteristic makes CsI widely applicable as a fluorescent agent in areas like chemical analysis, physical experimentation, and medical fields \citep{rodney1955optical,baumeler2023surface,buaban2023calcium}.
\textbf{Lead Telluride (PbTe)} ranks as a material with the second-highest uncertainty among those with five polymorphic structures.
PbTe is widely used in thermoelectric devices, such as generators and coolers, due to its high thermoelectric efficiency \citep{heremans2008enhancement,wang2023fine}. 
Additionally, as a semiconductor with a narrow band gap, PbTe is highly responsive to infrared (IR) light, making it a valuable component in infrared detectors and sensors \citep{wojciechowski2020highly,bafekry2020electronic}.

In conclusion, uncertainty analysis highlights that \proposed~effectively captures the uncertainty related to the model prediction, the presence of polymorphic structures within a single composition, and the computational challenges associated with crystal structures, all of which are a common interest in materials science.
Moreover, we observe that the materials exhibiting high uncertainty often find diverse applications across numerous scientific domains, highlighting the practical effectiveness of \proposed~in materials discovery.

\subsection{Qualitative Analysis}
\label{sec: Qualitative Analysis}

While our previous analysis in Section \ref{sec: Uncertainty Analysis} shows an increase in the number of possible structures generally leads to an increase in the uncertainty, we do observe some instances where \proposed~exhibits high uncertainty in non-polymorphic compositions and minimal uncertainty in polymorphic compositions.
First, in Figure~\ref{fig:case study} (a), we observe that the composition of $\text{Hg}\text{Cl}$ and $\text{CaS}$ exhibit high uncertainty, even though they only have one possible structure.
We attribute this phenomenon to the limited availability of {element combinations of Hg, Cl and Ca, S} in the MP dataset, 
which occurred due to several factors including the rarity of certain elements and the difficulty in synthesizing substances with specific combinations of elements~\citep{castor2006rare,jang2020structure}. 
On the other hand, we observe the learned distribution of $\text{ScN}$ and $\text{AgSO}_4$ collapsed to zero even though each of them has three possible polymorphic structures (Figure~\ref{fig:case study} (b)).
This behavior arises from the structural similarity among the polymorphic structures, where all three polymorphic structures of each composition fall within the same cubic and monoclinic structural system, respectively.
That is, while we observe that an increase in the number of polymorphic structures generally leads to an increase in the uncertainty in Figure~\ref{fig:uncertainty}, this observation highlights that \textit{\proposed~acquires detailed insights concerning polymorphic structures beyond mere quantitative counts.}


\section{Discussion}

\subsection{Application Areas}
\label{App: Application Areas}

The practical applications of \fullname (\proposed) span a wide range of industries and research areas within materials science. 
By addressing the limitations posed by detailed structural data requirements and polymorphism, \proposed~offers a practical solution that can be leveraged in various contexts. 
For example, in catalyst development, \proposed~can accelerate the discovery of new catalysts by predicting the activity and selectivity of materials based on their elemental composition alone. This can help in identifying promising candidates for catalytic reactions without the need for extensive experimental validation of each structure. In the development of batteries, \proposed~can assist in predicting the electrochemical properties of electrode materials, such as lithium-ion batteries or solid-state batteries. This can streamline the process of finding materials with high energy density and stability. For semiconductor devices, selecting materials with optimal electronic properties is crucial. \proposed~can help predict the electrical conductivity, bandgap, and other properties of materials based on their composition, aiding in the selection of materials for next-generation electronics. In the pharmaceutical industry, \proposed~can be used to predict the solubility, stability, and bioavailability of drug compounds based on their elemental composition. This can speed up the drug discovery process by identifying viable drug candidates early in the pipeline.

\subsection{Broad Impact}
The introduction of Probabilistic Composition Representation Learning (PCRL) represents a significant advancement in the application of machine learning (ML) to materials science. The broad impact of this work can be understood through several key dimensions: 

First, the ability of \proposed~to learn compositional representations of material opens up new avenues for material discovery.
Once the compositional encoder is pre-trained, it can be used to predict any properties of interest using only compositional information, without the need for costly structural calculations. 
By expanding the range of materials and properties that can be studied using ML techniques, \proposed~has the potential to democratize access to ML-driven insights in materials science, especially for researchers who may lack the resources or expertise to obtain detailed structural data.

Moreover, as demonstrated in Section \ref{sec: Empirical Results}, the learned representations in \proposed~are highly transferable to a variety of downstream tasks. 
This high degree of transferability sets \proposed~apart from previous compositional representation learning approaches that typically learn target-specific representations. 
This characteristic is particularly advantageous in the field of materials science, where wet-lab experimental data are often scarce due to the high costs associated with experiments. 
Consequently, \proposed~can be successfully applied to predict properties even in cases with limited data, extending its utility and impact in materials research.

Finally, \proposed~offers a novel approach to material discovery by capturing the uncertainties associated with polymorphism.
Polymorphism is a common phenomenon in materials science, where a single composition can exist in multiple structural forms. 
By capturing the diversity of these forms probabilistically, \proposed~provides a deep understanding of material behavior that can guide the design and synthesis of new materials, which has been extensively demonstrated in Section \ref{sec: Uncertainty Analysis}. 
This is particularly relevant for applications such as catalyst design, battery materials, and pharmaceuticals, where subtle changes in structure can have dramatic effects on performance.

\subsection{Theoretical Uncertainty Analysis}

Uncertainties in ML models can be categorized into aleatoric uncertainty and epistemic uncertainty. 
Aleatoric uncertainty refers to the inherent variability or randomness present in the data generation process, whereas epistemic uncertainty pertains to a lack of knowledge or information about the best model or the true underlying process \citep{hullermeier2021aleatoric,lu2024uncertainty,chen2024uncertainty}. 
In this section, we analyze uncertainty within the framework of aleatoric and epistemic uncertainty.

\smallskip
\noindent \textbf{Aleatoric uncertainty.}
Crystals with identical chemical compositions can exist in various structural forms, known as polymorphs, under different physical conditions such as temperature, pressure, and synthesis methods.
These different structural forms introduce inherent uncertainty into the composition data itself. 
Additionally, measurement errors or fluctuations that occur during the measurement or synthesizing of crystal structures contribute to aleatoric uncertainty.

\smallskip
\noindent \textbf{Epistemic uncertainty.}
While we model the probability distribution of composition as a Gaussian distribution, the particles that constitute a material are distributed discretely~\citep{kohn1996density}. Consequently, the material itself follows a distinct probability distribution, which is a composite of the discrete distributions of its constituent particles~\citep{cousin2023exploiting}. 
In other words, the simplicity of the Gaussian distribution introduces epistemic uncertainty by not accurately replicating the distribution of polymorphs. 
Additionally, most databases are highly biased due to the costly data collection process \citep{jia2019anthropogenic}, which means the database used may not capture the entire distribution of polymorphs. This limitation can also lead the model to exhibit high epistemic uncertainty in regions where there is insufficient data for a specific polymorph.

\subsection{Related Works}

\smallskip
\noindent \textbf{Compositional Representation Learning.}
Recently, there has been significant research into graphical representations of materials, which often necessitate costly and sometimes impractical DFT calculations of atomic positions.
This constraint hinders their use in real-world materials discovery, where frequent atomic-level reorganization necessitates structural calculations throughout the process.
Fortunately, material representations can be alternatively constructed solely based on composition, which indicates the concentration of the constituent elements, without any knowledge of the crystal structure~\citep{damewood2023representations}.
While composition has historically played a role in effective materials design~\citep{callister1964materials,pauling1929principles}, it has been recently demonstrated that deep neural networks (DNNs) tend to outperform conventional approaches when large datasets are available.
Specifically, ElemNet~\citep{jha2018elemnet} takes elemental compositions as inputs and trains DNNs with extensive high-throughput OQMD dataset~\citep{kirklin2015open}, showing improvements in performance as the network depth increases, up to a point where it reaches 17 layers.
Roost~\citep{goodall2020predicting} utilizes GNNs for compositional representation learning by creating a fully connected graph in which nodes represent elements, allowing for the modeling of interactions between these elements.
Instead of the message-passing scheme, CrabNet~\citep{wang2021compositionally} introduces a self-attention mechanism to adaptively learn the representation of individual elements based on their chemical environment.
While these methods are trained for a specific task, \proposed~learn generalized compositional representations that are universally applicable for various tasks considering 1) the structural information and 2) polymorphism in crystal, both of which have not been explored before.

\smallskip
\noindent\textbf{Machine Learning for Materials.}
Recently, ML approaches have become game changers in the field of materials science, where traditional research has heavily relied on theory, experimentation, and computer simulation, which is costly~\citep{wei2019machine,zhong2022explainable,zhang2023artificial}.
Among various ML methods, graph neural networks (GNNs) have been rapidly adopted by modeling crystal structures as graphical descriptions inspired by the recent success of GNNs in biochemistry~\citep{gilmer2017neural,stokes2020deep,jiang2021could}.
Specifically, CGCNN~\citep{xie2018crystal} first proposes a message-passing framework based on a multi-edge graph to capture interactions across cell boundaries, resulting in highly accurate prediction for eight distinct material properties. 
Building upon this multi-edge graph foundation, MEGNet~\citep{chen2019graph} predicts various crystal properties by incorporating a physically intuitive strategy to unify multiple GNN models.
Moreover, ALIGNN~\citep{choudhary2021atomistic} proposes to utilize a line graph, in addition to a multi-edge graph, to model additional structural features such as bond angles and local geometric distortions.
Despite the recent success of graph-based approaches, their major restriction is the requirement of atomic positions, 
which are typically determined through computationally intensive and sometimes infeasible DFT calculations.
As a result, their effectiveness is mainly demonstrated in predicting properties for systems that have already undergone significant computational effort, restricting their utility in the materials discovery workflow~\citep{damewood2023representations}.
We provide additional related works regarding the probabilistic representation learning and crystal structure prediction in Appendix \ref{App: Additional Related Works}.


\subsection{Limitations and Future Works}

\noindent \textbf{Pre-Training Dataset.}
While we train the model using the MP database, it is important to note that the structures in the MP database predominantly originate from experimental results accumulated over several decades, encompassing a wide range of varying conditions. MP aggregates these structures and relaxes them through energy minimization at T=0 and P=0. This relaxation can either result in significant structural transformations into a more stable polymorph at T=0 and P=0 or cause minor adjustments if the initial experimental structure is already near a local minimum.

Although the polymorphs used in our work may not precisely represent conditions at varying temperatures and pressures, we believe they still capture the relevant physical trends, as the structural data is consistently computed under standard conditions of 0K and 0 atm. Furthermore, numerous prior studies have shown the effectiveness of pretraining models on DFT-calculated properties at 0K and 0 atm for predicting a wide range of experimental properties observed in wet lab environments \cite{jha2019enhancing}.
For future work, it would be valuable to explore the collection of initial crystal structures under consistent conditions, though this will require additional effort.

\smallskip
\noindent \textbf{Appropriateness of Gaussian Assumption.}
In this paper, we choose the Gaussian distribution as a prior, which has multiple advantages when incorporated with deep neural networks as follows:
\begin{itemize}
    \item Efficient gradient computation is available with a reparameterization trick.
    \item Analytical computation of KL divergence is available and theoretically guaranteed.
\end{itemize}
In fact, the particles that make up a material are distributed discretely~\citep{kohn1996density}, and the material itself manifests following a distinct probability distribution, which is a composite of the discrete distributions attributed to its constituent particles~\citep{cousin2023exploiting}. 
While parameterizing such discrete distributions for each material is nearly impossible, it would be a promising direction for future work to explore more complex prior distributions that can be parameterized by neural networks.

\subsection{Conclusion}

This paper focuses on learning a probabilistic representation of composition that incorporates polymorphism of crystalline materials.
Given composition and its corresponding polymorphic structures, \proposed~learns a parameterized Gaussian distribution for each composition, whose mean becomes the representation of composition and variance indicates the level of uncertainty stemming from the polymorphic structures.
Extensive empirical studies on sixteen datasets have been conducted to validate the effectiveness of \proposed~in learning compositional representations.
Moreover, a comprehensive analysis of uncertainties reveals that the model learns diverse complexities encountered in materials science, highlighting the practicality of \proposed~in real-world material discovery process.

\paragraph{Intended Users}
\proposed~is intended for chemistry, material science, and artificial intelligence researchers and data scientists who want to apply AI algorithms and innovate novel methods to tackle problems formulated in \proposed~datasets and tasks.

\paragraph{Code Availability} The source code for \proposed~is available at~\url{https://github.com/Namkyeong/PCRL}.

\paragraph{Computing Resources} 
We perform all pre-training and downstream tasks on a 24GB NVIDIA GeForce RTX 3090 GPU.

\paragraph{Ethics Statement}
The development and dissemination of the \proposed~adhere to stringent ethical standards to ensure the responsible use of the data. 
The sources of the data are clearly documented in Section \ref{app:datasets}. 
Since this paper focuses on material discovery, we believe there are no additional ethical concerns to declare.

\paragraph{Competing Interests} The authors declare no competing interests. 

\clearpage

\rule[0pt]{\columnwidth}{3pt}
\begin{center}
    \Large{\bf{{Supplementary Material for \\ Compositional Representation of Polymorphic Crystalline Materials}}}
\end{center}
\vspace{2ex}

\DoToC

\clearpage

\appendix

\section{Structural Graph Construction}
\label{App: Structural Graph Construction}
In this section, we provide the detailed structural graph construction process with a figure.
Overall, this structural graph is the same as previous works~\citep{xie2018crystal,yan2022periodic}.
Given a crystal structure $(\mathbf{P}, \mathbf{L})$, suppose the unit cell has $n_s$ atoms, we have $\mathbf{P} = {[\mathbf{p}_1, \mathbf{p}_2, \ldots, \mathbf{p}_{n_s}]}^\intercal \in \mathbb{R}^{{n_s} \times 3}$ indicating the atom position matrix and $\mathbf{L} = {[\mathbf{l}_1, \mathbf{l}_2, \mathbf{l}_3]}^\intercal \in \mathbb{R}^{3 \times 3}$ representing the lattice parameter describing how a unit cell repeats itself in three directions.
Since the crystal usually possesses irregular shapes in practice, $\mathbf{l}_1, \mathbf{l}_2, \mathbf{l}_3$ are not always orthogonal in 3D space~\citep{yan2022periodic}.
For clear visualization, we provide examples of periodic patterns in 2D space in Figure~\ref{fig: structural graph construction} (a).

Based on the crystal parameters $(\mathbf{P}, \mathbf{L})$, we construct a multi-edge graph $\mathcal{G}^b = (\mathcal{V}, \mathbf{A}^b)$ that captures atom interactions across cell boundaries~\citep{xie2018crystal}.
Specifically, $v_i \in \mathcal{V}$ denotes an atom $i$ and all its duplicates in the infinite 3D space whose positions are included in the set $\{\hat{\mathbf{p}}_i | \hat{\mathbf{p}}_i = \mathbf{p}_i+k_1\mathbf{l}_1+k_2\mathbf{l}_2+k_3\mathbf{l}_3, k_1, k_2, k_3 \in \mathbb{Z} \}$, where $\mathbb{Z}$ denotes the set of all the integers.
Moreover, $\mathbf{A}^b \in \{0,1\}^{n_s\times n_s}$ denotes an adjacency matrix, where $\mathbf{A}_{i,j}^b = 1$ if two atoms $i$ and $j$ are within the predefined radius $r$ and $\mathbf{A}^{b}_{ij} = 0$ otherwise.
Specifically, nodes $v_i$ and $v_j$ are connected if there exists any combination $k_1, k_2, k_3 \in \mathbb{Z}$ such that the euclidean distance $d_{ij}$ satisfies $d_{ij} = {\|\mathbf{p}_i + k_1\mathbf{l}_1 + k_2\mathbf{l}_2 + k_3\mathbf{l}_3 - \mathbf{p}_j \|}_2 \leq r$ (see Figure~\ref{fig: structural graph construction} (b)).
For the initial feature for edges, we expand the distance $d_{ij}$ between atom $v_i$ and $v_j$ by Gaussian basis following previous works~\citep{xie2018crystal}.
Moreover, each node in $\mathcal{G}^b$ is associated with a learnable feature $\mathbf{x}^b \in \mathbb{R}^{F}$, which is shared across all crystals, to make sure we utilize only structural information.

\begin{figure}[h]
    \centering
    \includegraphics[width=0.6\linewidth]{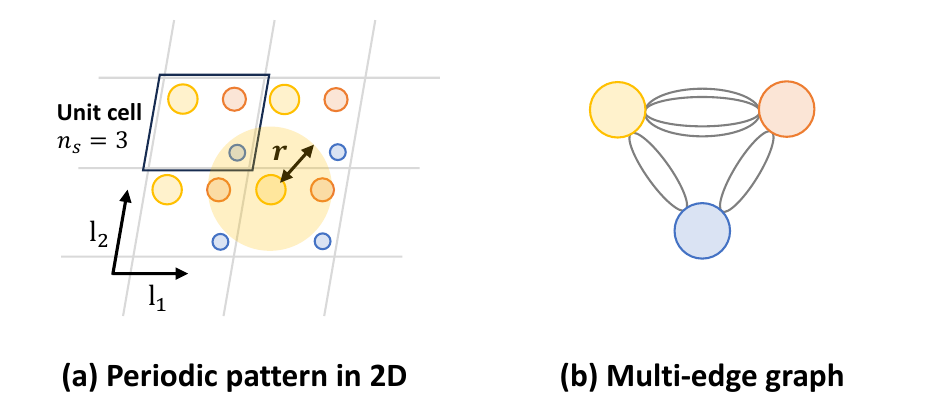}
    \caption{Structural graph construction.}
    \label{fig: structural graph construction}
\end{figure}

\section{Implementation Details}
\label{App: Implementation Details}

In this section, we provide implementation details of \proposed.

\subsection{Structural Graph Encoder}
\label{App: Structural Graph Encoder}
Our structural graph encoder comprises two components: the encoder and the processor. 
The encoder acquires the initial representation of atoms and bonds, while the processor is responsible for learning how to pass messages throughout the crystal structure.
More formally, given an atom $v_i$ and the bond $e_{ij}$ between atom $v_i$ and $v_j$ in crystal structure, node encoder $\phi_{node}$ and edge encoder $\phi_{edge}$ outputs initial representations of atom $v_i$ and bond $e_{ij}$ as follows:
\begin{equation}
    \mathbf{h}^{0, b}_{i} = \phi_{node}(\mathbf{X}^{b}),~~~ \mathbf{b}^{0, b}_{ij} = \phi_{edge}(\mathbf{B}^b_{ij}),
\end{equation}
where $\mathbf{X}^b \in \mathbb{R}^{n_{s} \times F}$ is the atom feature matrix whose $i$-th row indicates the input feature of atom $v_i$, $\mathbf{B}^b \in \mathbb{R}^{n_s \times n_s \times F} $ is the bond feature tensor.
As previously explained in Section~\ref{sec: Structural Graph Construction}, we employ a common $\mathbf{x}^b$ for all atoms across all crystals, resulting in every row in $\mathbf{X}^b$ being identical to $\mathbf{x}^b$.
With the initial representations of atoms and bonds, the processor learns to pass messages across the crystal structure and update atom and bond representations as follows:
\begin{equation}
    \mathbf{b}^{l+1, b}_{ij} = \psi^{l}_{edge}(\mathbf{h}^{l, b}_{i}, \mathbf{h}^{l, b}_{j}, \mathbf{b}^{l, b}_{ij}), ~~~ 
    \mathbf{h}^{l+1, b}_{i} = \psi^{l}_{node}(\mathbf{h}^{l, b}_{i}, \sum_{j \in \mathcal{N}(i)}{\mathbf{b}^{l+1, b}_{ij}}),
\end{equation}
where $\mathcal{N}(i)$ is the neighboring atoms of atom $v_i$, $\psi$ is a two-layer MLP with non-linearity, and $l = 0, \ldots, L'$.
Note that $\mathbf{h}^{L', b}_{i}$ is equivalent to the $i$-th row of the atom embedding matrix $\mathbf{Z}^b$ in Equation~\ref{eq: structural}.
In this paper, we use a 3-layered structural graph encoder, i.e., $L' = 3$.

\subsection{Probabilistic Composition Encoder}
\label{App: Probabilistic Composition Encoder}

\textbf{Composition Graph Encoder $f^a$.} For the composition graph encoder $f^a$, we utilize the architecture of GCNs~\citep{kipf2016semi} and Jumping Knowledge Network~\citep{xu2018representation}.
Specifically, given elemental feature matrix $\mathbf{X}^a$ and adjacency $\mathbf{A}^a$, GCN layers pass the messages to obtain latent elemental feature matrix as follows:
\begin{equation}
    \mathbf{h}^{l+1, a}_{i} = \text{GCN}^{l}(\mathbf{h}^{l, a}_{i}, \mathbf{A}^a),
\end{equation}
where $\mathbf{h}^{0, a}_{i}$ indicates $i$-th row of elemental feature matrix $\mathbf{X}^a$, and $l = 0, \ldots, L'$.
After $L'$ step message passing steps, we obtain a final representation of composition as follows:
\begin{equation}
    \mathbf{Z}^{a}_{i} = \mathbf{W}(\text{Concat}[\mathbf{h}^{0, a}_{i}, \cdots, \mathbf{h}^{L', a}_{i}]),
\end{equation}
where $\mathbf{W} \in \mathbb{R}^{F \times L'F}$ is a learnable weight matrix that reduces the dimension of concatenated representations. Note that $\mathbf{Z}^{a}_{i}$ is equivalent to the $i$-th row of the element embedding matrix $\mathbf{Z}^{a}$ in Equation~\ref{eq: composition}.
We also use $L' = 3$ for composition encoder $f^a$.
After obtaining the elemental representation matrix $\mathbf{Z}^a$, we obtain compositional representation $\mathbf{z}^a$ by employing weighted sum pooling, which takes into account the compositional ratio.

\textbf{Mean $f_{\mu}^a$ and Variance $f_{\sigma}^a$ Module.}
After obtaining the elemental representation matrix $\mathbf{Z}^a$, we utilize set2set~\citep{vinyals2015order} pooling function to obtain the mean vector and diagonal entries of the covariance vector.
More specifically, given $\mathbf{Z}^a$, we obtain mean vector $\mathbf{z}^{a}_{\mu}$ and diagonal covariance vector $\mathbf{z}^{a}_{\sigma}$ as follows:
\begin{equation}
    \mathbf{z}^{a}_{\mu} = \mathbf{\hat{z}}^{a}_{\mu} + \mathbf{z}^{a}, ~~~ \mathbf{\hat{z}}^{a}_{\mu} = \text{Set2set}_{\mu}(\mathbf{Z}^{a}),
\end{equation}
\begin{equation}
    \mathbf{z}^{a}_{\sigma} = \mathbf{\hat{z}}^{a}_{\sigma} + \mathbf{z}^{a}, ~~~ \mathbf{\hat{z}}^{a}_{\sigma} = \text{Set2set}_{\sigma}(\mathbf{Z}^{a}).
\end{equation}
By obtaining mean and diagonal covariance vectors with separate pooling functions, i.e., $\text{Set2set}_{\mu}$ and $\text{Set2set}_{\sigma}$, the model learns different attentive aspects involved for each module.

\subsection{Training Details}
\label{App: Training Details}

We also describe the implementation details to enhance the reproducibility. 
Our method is implemented on Python 3.7.1, PyTorch 1.8.1, and Torch-geometric 1.7.0.
All experiments are conducted using a 24GB NVIDIA GeForce RTX 3090.
Model hyperparameters are given in Table~\ref{tab: hyperparameters}.
During training, we clip the gradient to the maximum value of 2 for stability~\citep{zhang2019gradient}.

\begin{table}[h]
\caption{Hyperparameter specifications of \proposed.}    
    \centering
    \resizebox{0.7\linewidth}{!}{
    \begin{tabular}{cccccccccccc}
    \toprule
    \multicolumn{2}{c}{$\#$ Layers} & & Hidden & Learning & Batch & \multirow{2}{*}{Epochs} & Number of & \multirow{2}{*}{$\beta$} & & \multicolumn{2}{c}{Initial} \\
    \cmidrule{1-2} \cmidrule{11-12}
    $f^a$ & $f^b$ &  & dim ($F$) & Rate ($\eta$) & Size &  & Samples ($J$) &  & & $c$ & $d$ \\
    \midrule
    \midrule
    3 & 3 & & 200 & 5e-05 & 256 & 100 & 8 & 1e-08 & & 20 & 20\\
    \bottomrule
    \end{tabular}}
    \label{tab: hyperparameters}
\end{table}

\section{Additional Experiments}
\label{App:additional Experiments}



\subsection{Representation Space Analysis}
\label{App: Representation Space Analysis}
\textbf{Alignment of Composition and Polymorphic Structures.}
In this part of our study, we delve deeper into the analysis of the representation space for composition and its related polymorphic structures. 
For a clearer understanding, we've utilized t-SNE to visualize the representations of composition that correspond to 8 and 10 polymorphic structures.
In the visualizations, each composition is marked by a circle, while the polymorphic structures are denoted by 'x' markers.
In Figure~\ref{fig: app representation space}, we observe that the composition and its corresponding polymorphic structures are located close to each other, indicating that our composition encoder has effectively learned the information of the polymorphic structures.

\begin{figure}[h]
    \centering
    \includegraphics[width=0.7\linewidth]{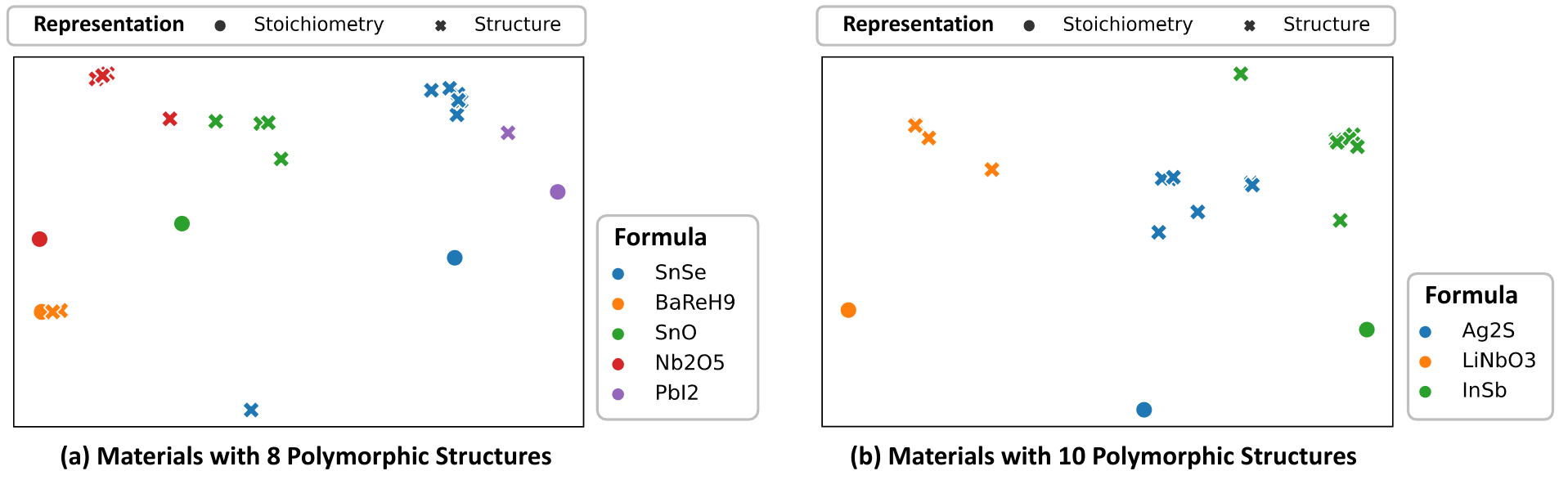}
    \caption{Representation space of composition and its corresponding polymorphic structures.}
    \label{fig: app representation space}
\end{figure}

\textbf{Compositional Representation with Multiple Aspects.}
In this section, we investigate the various aspects of learned compositional representation.
To do so, we first visualize the learned representation of composition in Band Gap dataset using t-SNE \cite{van2008visualizing} in Figure~\ref{fig:representation space}.
Although \proposed~focusing solely on structural information to learn compositional representation, we find that the representation encompasses multiple aspects, 
including not just the material properties such as band gap (Figure~\ref{fig:representation space} (a)), but also the uncertainties arising from polymorphism (Figure~\ref{fig:representation space} (b)).
This implies that \proposed~can reliably provide insights into the uncertainties related to polymorphism derived from specific composition, emphasizing its valuable application in material discovery processes.

\begin{figure}[h]
\centering
    \includegraphics[width=0.7\linewidth]{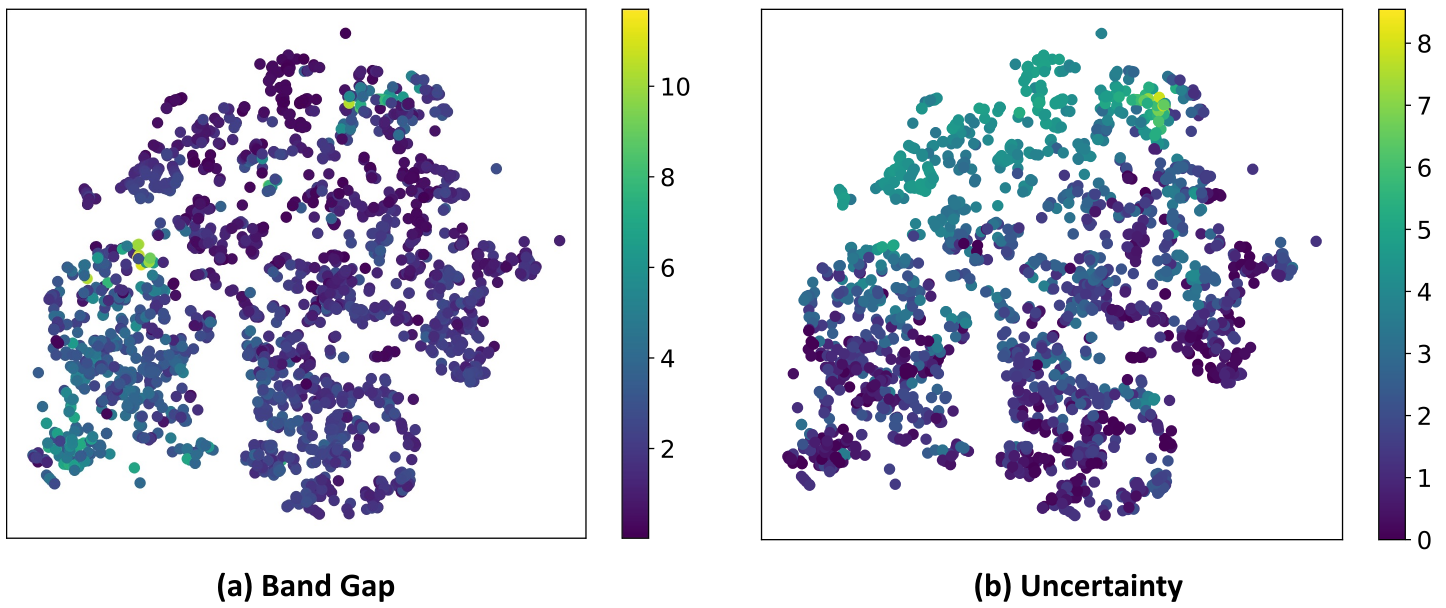}
    \caption{Visualization of compositional representations in Band Gap dataset.}
    \label{fig:representation space}
\end{figure}

\subsection{Statistical Significance Test}
\label{App: Statistical Significance Test}

To exhibit the statistical significance of the enhancement, we conducted a paired t-test within the representation learning scenario for each dataset (Table~\ref{tab: main table}). 
To do so, we additionally conducted 5 separate instances of 5-fold cross-validation, with each instance featuring entirely distinct data splits (amounting to a total of 25 individual runs), and then comparing the means from each set of cross-validation results. The resulting numbers in table represent the p-values from the paired t-tests conducted with \proposed~under the following hypotheses:
$H_{0}: \mu_{PCRL} = \mu_{Baseline}$, and
$H_{1}: \mu_{PCRL} > \mu_{Baseline}$.

\begin{table}[h]
\caption{Statistical significance test results.}
    \centering
    \resizebox{0.95\linewidth}{!}{
    \begin{tabular}{lccccccccccc}
    \toprule
    \multirow{3}{*}{\textbf{Model}} & \multirow{3}{*}{\textbf{Band G.}} & \multirow{3}{*}{\textbf{Form.~E.}} & \multirow{3}{*}{\textbf{Metallic}} & &\multicolumn{3}{c}{\textbf{ESTM 300K}} & &\multicolumn{3}{c}{\textbf{ESTM 600K}} \\
    \cmidrule{6-8} \cmidrule{10-12}
    & & & & & E.C. & T.C. & Seebeck &  & E.C. & T.C. & Seebeck \\
    \midrule
    GraphCL & $3.80e^{-4}$ & $4.18e^{-3}$ & $7.63e^{-3}$ & & $3.12e^{-3}$ & $4.29e^{-4}$ & $2.83e^{-3}$ & & $6.72e^{-4}$ & $1.39e^{-2}$ & $4.07e^{-2}$ \\
    MP Band G. & $2.48e^{-1}$ & $1.20e^{-4}$ & $6.53e^{-4}$ & & $4.48e^{-4}$ & $2.15e^{-4}$ & $4.55e^{-4}$  & & $5.90e^{-4}$ & $1.47e^{-3}$  & $1.40e^{-3}$  \\
    MP Form. E. & $5.14e^{-2}$  & $9.23e^{-1}$ & $1.89e^{-2}$ &  & $3.93e^{-4}$  & $1.75e^{-3}$ & $5.03e^{-3}$ & & $1.47e^{-3}$  & $3.23e^{-3}$  & $4.30e^{-2}$  \\
    3D Infomax & $1.21e^{-3}$ & $1.40e^{-3}$ & $3.42e^{-2}$   &  & $1.44e^{-2}$  & $2.72e^{-2}$ & $8.39e^{-4}$ & & $6.06e^{-3}$  & $3.89e^{-2}$  & $5.64e^{-2}$  \\
    \bottomrule
    \end{tabular}}
    \label{tab: app statistical significance}
\end{table}

We have the following observations:
\textbf{1)} \proposed~achieves statistically significant improvements over baseline methods that do not leverage structural information (such as GraphCL, MP Band G., MP Form. E.) at the 0.05 significance level except for Band G. and Form. E. datasets.
\textbf{2)} On the other hand, in the datasets for Band G. and Form. E., \proposed~does not achieve statistical significance compared to models that benefit from the corresponding DFT-calculated values. For example, in Band G. dataset, we observe that the model trained with DFT-calculated band gap (i.e., MP Band G.) outperforms \proposed, which was also described in observation 3 in Section~\ref{sec: Empirical Results} in the main paper. However, as discussed in the main text, these models are specifically tailored for particular properties, which may limit their broader applicability across a range of tasks.

\subsection{Fine-tuning on Out-of-Distribution Scenarios}
\label{App: Fine-tuning OOD}

In this section, we assess model performance in out-of-distribution scenarios by dividing the evaluation dataset according to composition information. 
Although we randomly split the dataset and conducted 5-fold cross-validation in the main manuscript, for this particular experiment, the training set excludes elements from the Actinide and Lanthanide series, whereas the test set includes them \cite{katz1987chemistry,seaborg1993overview}.
The model parameters are fine-tuned with the training set and evaluated on the test set.
In Table~\ref{tab: fine tuning ood}, we observe that \proposed~surpasses baseline models in seven out of nine datasets, illustrating that the \proposed~training method is also effective in out-of-distribution scenarios, which is crucial in real-world materials discovery.

\begin{table}[h]
\caption{Fine-tuning performance on unseen types of materials (MAE).}
    \centering
    \resizebox{0.95\linewidth}{!}{
    \begin{tabular}{lcccccccccccccc}
    \toprule
    \multirow{3}{*}{\textbf{Model}} & \multicolumn{2}{c}{\textbf{DFT}} & \multirow{3}{*}{\textbf{Poly.}} & \multirow{3}{*}{\textbf{Band G.}} & \multirow{3}{*}{\textbf{Form.~E.}} & \multirow{3}{*}{\textbf{Metallic}} & &\multicolumn{3}{c}{\textbf{ESTM 300K}} & &\multicolumn{3}{c}{\textbf{ESTM 600K}} \\
    \cmidrule{2-3} \cmidrule{9-11} \cmidrule{13-15}
    & \textbf{Prop.}& \textbf{Str.} & & & &  & & E.C. & T.C. & Seebeck &  & E.C. & T.C. & Seebeck \\
    \midrule
    Rand init. & \xmark & \xmark & \xmark & 0.599  & 1.507  &0.259 & & 1.379 & 0.371 & 0.416 & & 0.832  & 0.275 & 0.866 \\
    GraphCL & \xmark & \xmark & \xmark & 0.589  & 1.531 & 0.253 & & 1.516 & \textbf{0.312} & \textbf{0.382} & & 1.096 & 0.291 & 0.829 \\
    MP Band G. & \cmark & \xmark & \xmark & 0.563 & 1.239 & 0.262 & & 1.757 &0.421  & 0.556 & & 0.836 & 0.302 & 0.975 \\
    MP Form.~E. & \cmark & \xmark & \xmark & 0.570 & 1.318 & 0.259 & & 1.580 & 0.397 & 0.454 & & 0.993 & 0.247 & 0.780 \\
    3D Infomax & \xmark & \cmark & \xmark & 0.562  & 1.343 &0.250 & &1.668  & 0.355 & 0.399  & & 0.853 & 0.279 & 0.858 \\
    \midrule
    \proposed & \xmark & \cmark & \cmark & \textbf{0.546}  & \textbf{1.186}  & \textbf{0.244} & & \textbf{1.320} & 0.412  & 0.403 & & \textbf{0.794} & \textbf{0.236} & \textbf{0.729} \\
    \bottomrule
    \end{tabular}}
    \label{tab: fine tuning ood}
\end{table}

\subsection{Model Analysis}
\label{App: Model Analysis}

\textbf{Sensitivity Analysis.}
In addition to model analysis in Section~\ref{sec: Empirical Results}, we provide an analysis on various hyperparameters in \proposed, i.e., initial values of $c, d$ and number of samples $J$ in Equation~\ref{eq: soft con}.
We have the following observations:
\textbf{1)} While we made $c$ and $d$ learnable parameters to allow the model to adjust them adaptively to an optimal point, we've also found that setting the initial values for $c$ and $d$ is crucial in model training. 
This indicates that initial value plays a significant role in guiding the model correctly from the outset of the training process, ultimately contributing to good performance.
\textbf{2)} On the other hand, we observe \proposed~shows robustness in various numbers of samples, suggesting that it can be trained effectively without a large number of samples, which will demand an extensive amount of computational resources.

\begin{figure}[h]
    \centering
    \includegraphics[width=0.6\linewidth]{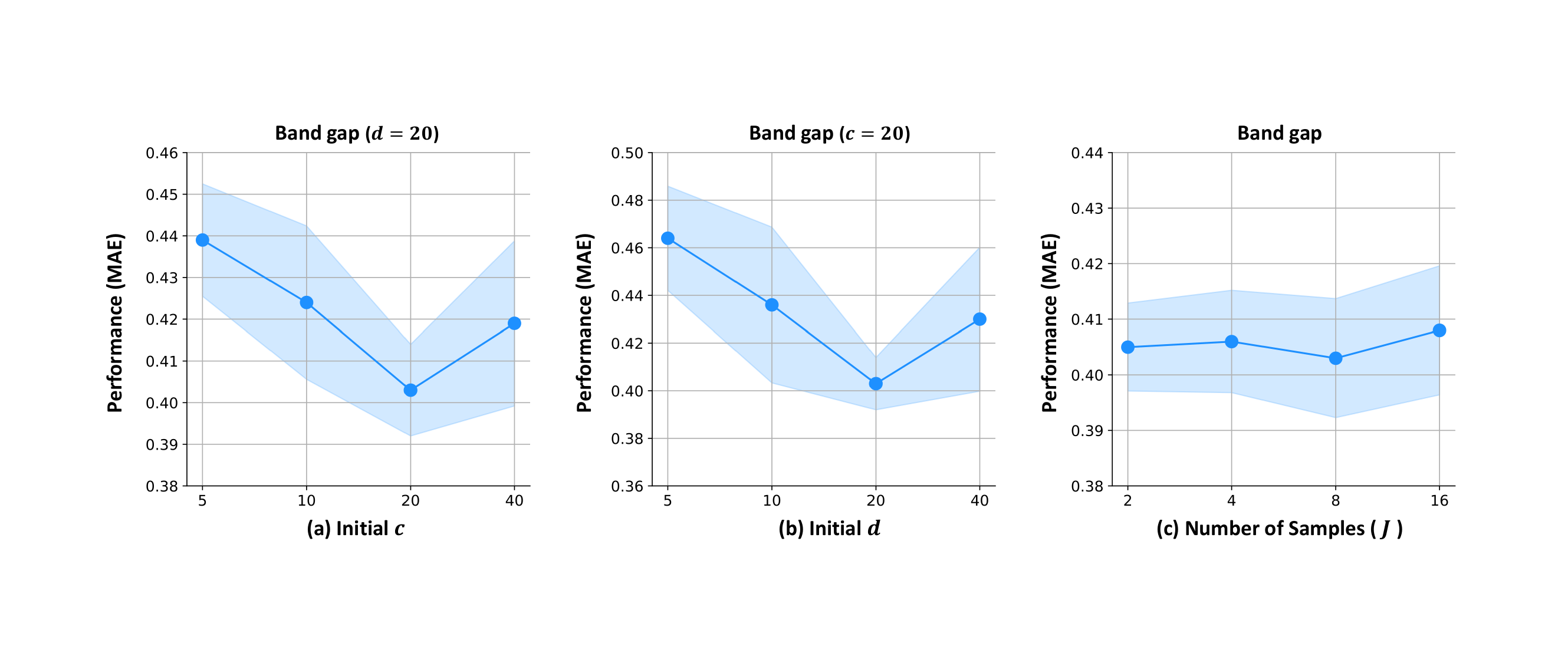}
    \caption{Additional sensitivity analysis results.}
    \label{fig: additional sensitivity}
\end{figure}

\noindent \textbf{Ablation Studies.}
In this section, we conduct ablation studies on our model by removing the sampling process described in Equation~\ref{eq: mu_sigma}, which is denoted as "w/o Sampling" in Table~\ref{tab: ablation studies}. 
To clarify, rather than utilizing the sampled representations $\hat{\textbf{z}}_j^a$ in Equation~\ref{eq: match prob}, we directly employ the mean vector of composition, denoted as $\textbf{z}_{\mu}^a$, for the soft contrastive loss.
By doing so, the model transitions from learning a probabilistic representation of composition to learning a deterministic representation of composition.
To compare with methods that don't incorporate polymorphic structural information, 
such as {3D Infomax}, we also present the performance of {3D Infomax} in Table~\ref{tab: ablation studies}.
We have the following observations:
\textbf{1)} Considering polymorphic structure is crucial in compositional representation learning by comparing {3D Infomax} and {w/o Sampling}.
\textbf{2)} Additionally, the sampling process typically leads to improved performance, underscoring the advantage of learning a probabilistic representation of composition.
While {w/o Sampling} outperforms \proposed~in two datasets, the absence of the sampling process means the model can no longer estimate uncertainty in composition, thereby losing its practicality in real-world materials discovery.
In summary, we argue that \proposed~learns a probabilistic compositional representation, which not only enables accurate uncertainty estimation but also enhances model performance.

\begin{table}[h]
\caption{Ablation studies in representation learning scenarios (MAE).}
    \centering
    \resizebox{0.8\linewidth}{!}{
    \begin{tabular}{lcccccccccccccc}
    \toprule
    \multirow{3}{*}{\textbf{Model}} & \multicolumn{2}{c}{\textbf{DFT}} & \multirow{3}{*}{\textbf{Poly.}} & \multirow{3}{*}{\textbf{Band G.}} & \multirow{3}{*}{\textbf{Form.~E.}} & \multirow{3}{*}{\textbf{Metallic}} & &\multicolumn{3}{c}{\textbf{ESTM 300K}} & &\multicolumn{3}{c}{\textbf{ESTM 600K}}\\
    \cmidrule{2-3} \cmidrule{9-11} \cmidrule{13-15}
    & \textbf{Prop.}& \textbf{Str.} & & & &  & & E.C. & T.C. & Seebeck &  & E.C. & T.C. & Seebeck\\
    \midrule
    \multirow{2}{*}{3D Infomax} & \multirow{2}{*}{\cmark} & \multirow{2}{*}{\cmark} & \multirow{2}{*}{\xmark} & 0.428  & 0.654 &0.201 & &0.969  & 0.217 &0.432  & & 0.692 & 0.212 & 0.428 \\
    & & & &\scriptsize{(0.015)}  &\scriptsize{(0.032)}  &\scriptsize{(0.032)}  & &\scriptsize{(0.110)}  &\scriptsize{(0.040)}  &\scriptsize{(0.070)} & &\scriptsize{(0.102)} &\scriptsize{(0.013)}  &\scriptsize{(0.076)} \\
    \multirow{2}{*}{w/o Sampling} & \multirow{2}{*}{\cmark} & \multirow{2}{*}{\cmark} & \multirow{2}{*}{\cmark} & 0.410 & 0.618 & 0.198 & & \textbf{0.864} & 0.208 & 0.407  & & 0.679 & 0.198  & \textbf{0.396}  \\
    & & & &\scriptsize{(0.006)}  &\scriptsize{(0.060)}  &\scriptsize{(0.030)} & &\scriptsize{(0.192)}  &\scriptsize{(0.027)}  &\scriptsize{(0.054)} & & \scriptsize{(0.084)} & \scriptsize{(0.011)}  &\scriptsize{(0.033)} \\
    \midrule
    \multirow{2}{*}{\proposed} & \multirow{2}{*}{\cmark} & \multirow{2}{*}{\cmark} & \multirow{2}{*}{\cmark} & \textbf{0.407} & \textbf{0.592}  & \textbf{0.194} & & 0.912 & \textbf{0.197}  & \textbf{0.388} & & \textbf{0.665} & \textbf{0.189} & 0.412 \\
    & & & &\scriptsize{(0.013)}  &\scriptsize{(0.039)}  &\scriptsize{(0.017)} &  &\scriptsize{(0.121)}  &\scriptsize{(0.020)}  &\scriptsize{(0.059)} & &\scriptsize{(0.126)}  &\scriptsize{(0.017)}  &\scriptsize{(0.043)} \\
    \bottomrule
    \end{tabular}}
    \label{tab: ablation studies}
\end{table}

\subsection{Pre-training on OQMD Dataset}
\label{App: Pre-training on OQMD Dataset}

In Table~\ref{app tab: oqmd dataset}, we present the comprehensive experimental results from Table \ref{tab: different database} in the main manuscript. 
We observe that the model trained with the MP database generally performs better than the one trained with the OQMD database.

\begin{table}[h]
\caption{Representation learning performance (MAE) comparison trained with different databases.}
    \centering
    \resizebox{0.8\linewidth}{!}{
    \begin{tabular}{lcccccccccccccc}
    \toprule
    \multirow{3}{*}{\textbf{Database}} & \multirow{3}{*}{\textbf{Band G.}} & \multirow{3}{*}{\textbf{Form.~E.}} & \multirow{3}{*}{\textbf{Metallic}} & &\multicolumn{3}{c}{\textbf{ESTM 300 K}} & &\multicolumn{3}{c}{\textbf{ESTM 600 K}} \\
    \cmidrule{6-8} \cmidrule{10-12}
    & & & & & E.C. & T.C. & Seebeck &  & E.C. & T.C. & Seebeck \\
    \midrule
    \multirow{2}{*}{MP (Ours)} & \textbf{0.407} & \textbf{0.592}  & \textbf{0.194} & & 0.912 & \textbf{0.197}  & \textbf{0.388} & & \textbf{0.665} & \textbf{0.189} & \textbf{0.412}\\
    &\scriptsize{(0.013)}  &\scriptsize{(0.039)}  &\scriptsize{(0.017)} &  &\scriptsize{(0.121)}  &\scriptsize{(0.020)}  &\scriptsize{(0.059)} & &\scriptsize{(0.126)}  &\scriptsize{(0.017)}  &\scriptsize{(0.043)} \\
    \midrule
    \multirow{2}{*}{OQMD} & 0.425  & 0.663  & \textbf{0.194} & & \textbf{0.814} & 0.216  & 0.401 & & 0.743 & 0.209 & 0.439\\
    &\scriptsize{(0.013)}  &\scriptsize{(0.039)}  &\scriptsize{(0.017)} &  &\scriptsize{(0.121)}  &\scriptsize{(0.020)}  &\scriptsize{(0.059)} & &\scriptsize{(0.126)}  &\scriptsize{(0.017)}  &\scriptsize{(0.043)} \\
    \bottomrule
    \end{tabular}}
\label{app tab: oqmd dataset}
\end{table}

\subsection{Many-to-many relationship between composition and structure}
\label{App: Many-to-many relationship}

In Table \ref{app tab: different relationship}, we present the comprehensive experimental results from Table \ref{tab: different relationship} in the main manuscript. 
It is evident that the model which focuses on learning the one-to-many relationship between composition and structure typically surpasses the performance of the model that addresses the many-to-many relationship.

\begin{table}[h]
\caption{Representation learning performance (MAE) comparison trained with different relationship.}
    \centering
    \resizebox{0.8\linewidth}{!}{
    \begin{tabular}{lcccccccccccccc}
    \toprule
    \multirow{3}{*}{\textbf{Relationship}} & \multirow{3}{*}{\textbf{Band G.}} & \multirow{3}{*}{\textbf{Form.~E.}} & \multirow{3}{*}{\textbf{Metallic}} & &\multicolumn{3}{c}{\textbf{ESTM 300 K}} & &\multicolumn{3}{c}{\textbf{ESTM 600 K}} \\
    \cmidrule{6-8} \cmidrule{10-12}
    & & & & & E.C. & T.C. & Seebeck &  & E.C. & T.C. & Seebeck \\
    \midrule
    \multirow{2}{*}{One-to-Many (Ours)} & \textbf{0.407} & \textbf{0.592}  & \textbf{0.194} & & 0.912 & \textbf{0.197}  & \textbf{0.388} & & \textbf{0.665} & \textbf{0.189} & \textbf{0.412}\\
    &\scriptsize{(0.013)}  &\scriptsize{(0.039)}  &\scriptsize{(0.017)} &  &\scriptsize{(0.121)}  &\scriptsize{(0.020)}  &\scriptsize{(0.059)} & &\scriptsize{(0.126)}  &\scriptsize{(0.017)}  &\scriptsize{(0.043)} \\
    \midrule
    \multirow{2}{*}{Many-to-Many} & 0.410  & 0.617  & 0.210 & & \textbf{0.892} & 0.199  & 0.416 & & 0.670 & 0.199 & 0.440 \\
    &\scriptsize{(0.012)}  &\scriptsize{(0.025)}  &\scriptsize{(0.023)} &  &\scriptsize{(0.117)}  &\scriptsize{(0.027)}  &\scriptsize{(0.070)} & &\scriptsize{(0.110)}  &\scriptsize{(0.053)}  &\scriptsize{(0.066)} \\
    \bottomrule
    \end{tabular}}
\label{app tab: different relationship}
\end{table}

\subsection{Case Studies}
\label{App: Case Studies}

From now on, we show case studies that are in line with our uncertainty analysis in Section \ref{sec: Uncertainty Analysis}.

\noindent \textbf{Low Uncertainty with Multiple Structures.}
In Figure~\ref{fig: app case study} (a), we observe two compositions (i.e., $\text{ZrC}$ and $\text{NdF}_3$) with collapsed uncertainty, even though they possess four distinct possible structures.
This phenomenon occurs because these structures share highly similar polymorphic arrangements, with only one unique structure in each composition. 
For instance, $\text{ZrC}$ and $\text{NdF}_3$ predominantly adopt cubic and hexagonal structures, respectively, with only one distinct possible structure for each composition.

\noindent \textbf{High Uncertainty with Multiple Structures.}
In this section, we present additional case studies that align with our expectations.
Figure~\ref{fig: app case study} (b) illustrates two compositions (i.e., $\text{NaI}$ and $\text{AlP}$) with the highest uncertainty among those possessing three polymorphic structures. 
For example, $\text{NaI}$ can exist in three distinct structures (i.e., cubic, orthorhombic, and tetragonal), and $\text{AlP}$ also exhibits three different structures (i.e., cubic, hexagonal, and tetragonal).
Given that varying atomic arrangements within materials lead to entirely distinct physical and chemical properties, it becomes crucial to convey the extent of structural diversity that composition can exhibit during the material discovery process. 
Therefore, these additional case studies highlight the practicality of \proposed~in real-world material discovery.

\begin{figure}[t]
    \centering
    \includegraphics[width=0.9\linewidth]{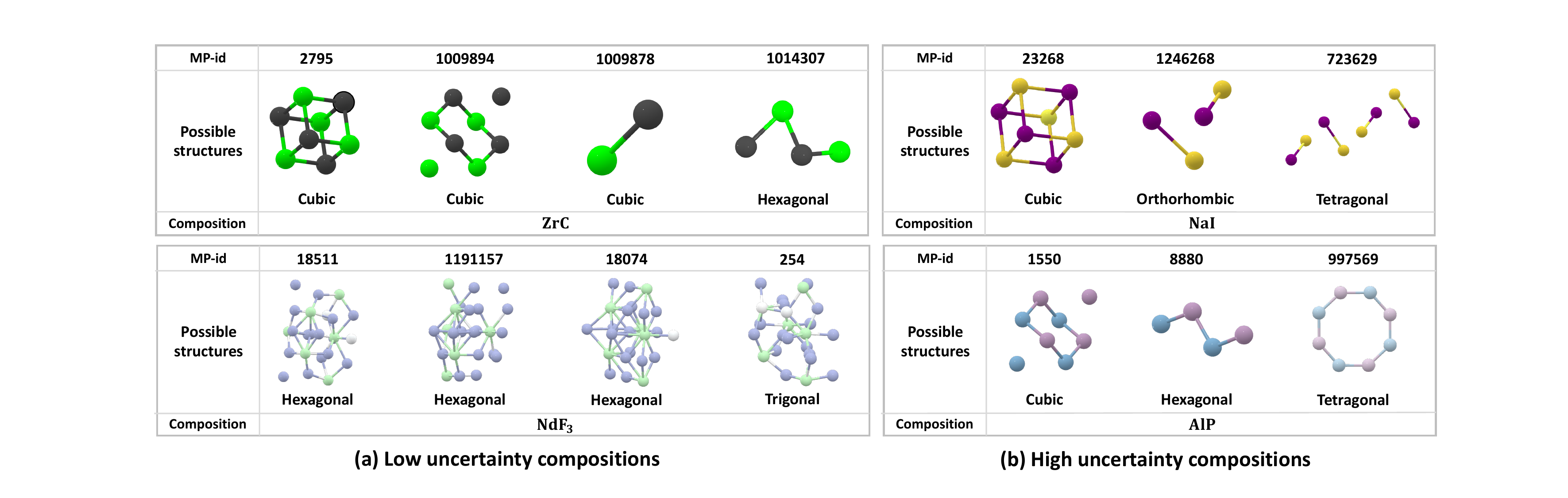}
    \caption{Additional qualitative uncertainty analysis.}
    \label{fig: app case study}
\end{figure}

\section{Related Works}
\label{App: Additional Related Works}

\subsection{Probabilistic Representation Learning}
First appearing in 2014 with the introduction of probabilistic word embeddings~\citep{vilnis2014word}, probabilistic representations got a surge of interest from ML researchers by offering numerous benefits in modeling uncertainty pertaining to a representation.
Specifically, in the computer vision domain, \citep{shi2019probabilistic} proposes to probabilistically represent face images to address feature ambiguity in real-world face recognition.
Moreover, \citep{oh2018modeling} introduces Hedged Instance Embeddings (HIB), which computes a match probability between point estimates but integrates it over the predicted distributions via Monte Carlo estimation.
This idea has been successfully extended to cross-modal retrieval~\citep{chun2021probabilistic}, video representation learning~\citep{park2022probabilistic}, and concept prediction~\citep{kim2023probabilistic}.
{
On the other hand, the importance of measuring uncertainties goes beyond the field of computer vision, particularly in the context of applying machine learning in the scientific discovery process~\citep{graff2021accelerating,wang2023scientific,zhang2023artificial}.
In our study, our objective is to develop a probabilistic representation of composition that incorporates uncertainties related to the different polymorphic forms associated with a single composition, thereby enhancing the reliability of the ML model for the material discovery process.
}

\subsection{Crystal Structure Prediction}

Crystal structure prediction (CSP) is the process of determining the stable three-dimensional structure of a compound from its chemical composition alone. Conventional CSP methods often combine density functional theory (DFT) with optimization algorithms. These algorithms carry out an iterative process searching for stable states that align with the local energy minima, while DFT is utilized to assess the energy at each step of the iteration~\citep{pickard2011ab,yamashita2018crystal,oganov2019structure}. 
A recent development in the field is DiffCSP~\citep{jiao2023crystal}, which employs a deep generative model within a diffusion framework to simultaneously optimize lattice matrices and atomic coordinates, offering a novel approach to the CSP problem.

It's important to highlight the growing interest among researchers in generating crystal structures without predefined composition, a direction distinct from CSP. The groundbreaking CDVAE~\citep{xie2021crystal} integrates a variational autoencoder (VAE) architecture with a diffusion-based decoder to produce the types of atoms, their coordinates, and lattice parameters. Unlike CSP-focused methods, CDVAE's primary goal is the generation of random crystal structures, providing a different avenue in the field of crystallography research.

\clearpage

\section{Notations}
\label{App: Notations}
In Table~\ref{table:notation}, we provide mathematical notations that are used in the main manuscript.

\begin{table}[h]
\centering
\caption{Mathematical notations. }
\resizebox{0.9 \columnwidth}{!}{
\begin{tabular}{c|p{0.74\columnwidth}}
\toprule[1pt]
Notations & Explanations \\ 
\midrule
$n_s$ & Number of atoms in crystal structure \\ 
$\mathbf{X}^b$ & An elemental feature matrix of structural graph \\ 
$\mathbf{A}^b$ & An adjacency matrix of structural graph \\ 
$\mathcal{G}^{b} = (\mathbf{X}^b, \mathbf{A}^b)$ & A crystal structural graph \\  
$\mathbf{z}^b$ & A latent representation of a crystal structural graph \\ 
$f^b$ & A GNN-based crystal structural encoder \\ 
\midrule
$n_e$ & Number of unique elements in a composition \\ 
$\mathcal{E} = \{e_1, \ldots, e_{n_e} \}$ & A unique set of elements in a composition\\ 
$\mathcal{R} = \{r_1, \ldots, r_{n_e} \}$ & A compositional ratio of each element in a composition \\
$\mathcal{G}^a = (\mathcal{E}, \mathcal{R}, \mathbf{A}^a)$ & A fully-connected composition graph \\ 
$\mathbf{X}^a$ & A elemental feature matrix of composition graph \\ 
$\mathbf{A}^a$ & An adjacency matrix of composition graph \\ 
$\Tilde{\mathbf{z}}^a$ & A sampled representation from latent distribution of composition \\ 
$f^a$ & A GNN-based composition graph encoder \\ 
$f_{\mu}^a$ & A mean module for composition graph \\ 
$f_{\sigma}^a$ & A variance module for composition graph \\ 
$\mathcal{P}^{\mathcal{G}^a}$ & A set of polymorphic structural graphs related to composition $\mathcal{G}^a$ \\
\midrule
$J$ & Number of samples from latent distribution of composition (Equation~\ref{eq: match prob}) \\
$c$ & Learnable parameters for scaling the Euclidean distance (Equation~\ref{eq: match prob}) \\
$d$ & Learnable parameters for shifting the Euclidean distance (Equation~\ref{eq: match prob}) \\
\midrule
$\mathcal{L}_{\text{con}}$ & Soft contrastive loss (Equation~\ref{eq: soft con}) \\ 
$\mathcal{L}_{\text{KL}}$ & KL divergence loss \\ 
$\beta$ & Hyperparameter that controls the weight of KL divergence loss \\ 
$\mathcal{L}_{\text{total}}$ & Total loss function (Equation~\ref{eq: total}) \\ 
\bottomrule[1pt]
\end{tabular}}
\label{table:notation}
\end{table}

\clearpage

\bibliographystyle{unsrt}
\bibliography{reference}

\end{document}